\documentclass[12pt]{article}

\usepackage[margin=1in]{geometry}

\RequirePackage[colorlinks,citecolor=blue,linkcolor=blue,urlcolor=blue,pagebackref]{hyperref}

\usepackage{microtype}
\usepackage{graphicx}
\usepackage{subfigure}
\usepackage{booktabs} 

\usepackage{hyperref}

\usepackage[utf8]{inputenc} 
\usepackage[T1]{fontenc}    
\usepackage{url}            
\usepackage{booktabs}       
\usepackage{amsfonts}       
\usepackage{nicefrac}       
\usepackage{microtype}      
\usepackage{xcolor}         

\usepackage{bm}
\usepackage{bbm}
\usepackage{comment}         
\usepackage{multicol}
\usepackage{multirow}
\usepackage{caption}
\usepackage{natbib}
\usepackage{braket}

\usepackage{amsmath,amssymb,amsthm}

\usepackage{color}

\usepackage{graphicx}

\DeclareMathOperator*{\argmin}{arg\,min}

\usepackage{times}

\newcommand{\pa}{\mathrm{\pa}}

\newcommand{\RN}[1]{%
  \textup{\uppercase\expandafter{\romannumeral#1}}%
}

\usepackage{xcolor}
\newcount\Comments  
\newcommand{\kibitz}[2]{\ifnum\Comments=1\textcolor{#1}{#2}\fi}

\newcommand{\mkato}[1]{\kibitz{black}{#1}}

\usepackage{algorithm}
\usepackage{algorithmic}

\usepackage{authblk}
\allowdisplaybreaks

\usepackage{amsmath}
\usepackage{amssymb}
\usepackage{mathtools}
\usepackage{amsthm}

\usepackage[capitalize,noabbrev]{cleveref}

\theoremstyle{plain}
\newtheorem{theorem}{Theorem}[section]
\newtheorem{proposition}[theorem]{Proposition}

\newtheorem{corollary}[theorem]{Corollary}
\theoremstyle{definition}
\newtheorem{definition}[theorem]{Definition}
\newtheorem{assumption}[theorem]{Assumption}
\theoremstyle{remark}
\newtheorem{remark}[theorem]{Remark}

\usepackage[textsize=tiny]{todonotes}

\title{Asymptotically Unbiased Synthetic Control Methods
by Moment Matching}

\usepackage{authblk}

\author{Masahiro Kato}
\author{Akari Ohda}

\affil{Department of Basic Science, the University of Tokyo}

\begin{document}

\maketitle

\begin{abstract}
{Synthetic Control Methods (SCMs) have become a fundamental tool for comparative case studies. The core idea behind SCMs is to estimate treatment effects by predicting counterfactual outcomes for a treated unit using a weighted combination of observed outcomes from untreated units. The accuracy of these predictions is crucial for evaluating the treatment effect of a policy intervention. Subsequent research has therefore focused on estimating SC weights. In this study, we highlight a key endogeneity issue in existing SCMs—namely, the correlation between the outcomes of untreated units and the error term of the synthetic control, which leads to bias in both counterfactual outcome prediction and treatment effect estimation. To address this issue, we propose a novel SCM based on moment matching, assuming that the outcome distribution of the treated unit can be approximated by a weighted mixture of the distributions of untreated units. Under this assumption, we estimate SC weights by matching the moments of the treated outcomes with the weighted sum of the moments of the untreated outcomes. Our method offers three advantages: first, under the mixture model assumption, our estimator is asymptotically unbiased; second, this asymptotic unbiasedness reduces the mean squared error in counterfactual predictions; and third, our method provides full distributions of the treatment effect rather than just expected values, thereby broadening the applicability of SCMs. Finally, we present experimental results that demonstrate the effectiveness of our approach.}
\end{abstract}

\section{Introduction}
\begin{figure}
    \centering
    \includegraphics[width=0.7\linewidth]{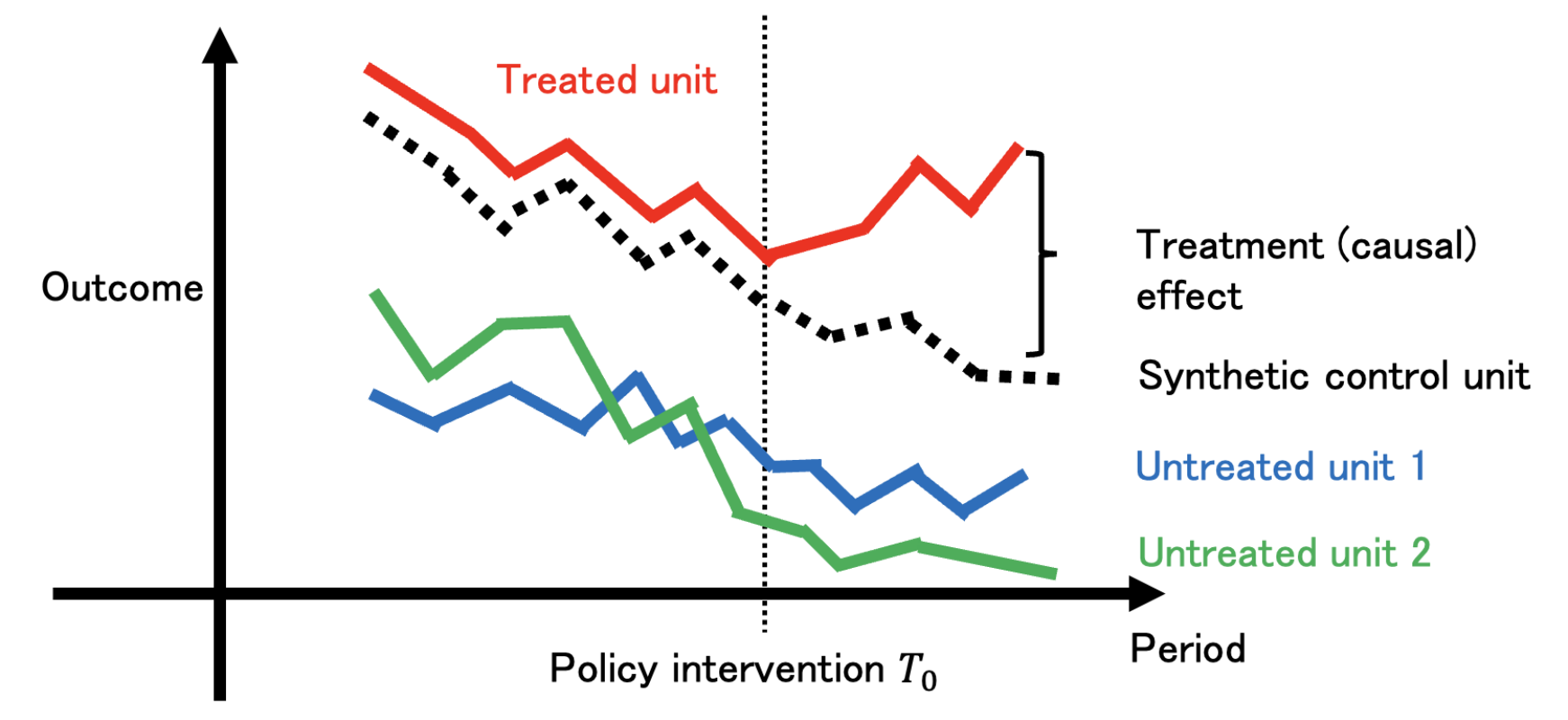}
    \caption{Illustration of SCMs}
    \label{fig:illustration}
\end{figure}

Synthetic control methods \citep[SCMs,][]{AbadieGardeazabal2003,AbadieDiamondHainmueller2010} have emerged as a crucial tool for program evaluation in comparative case studies across disciplines such as economics, statistics, and machine learning. In SCMs, there are multiple units, where one unit receives a policy intervention (treated unit) at a certain time while the remaining unitts remain untreated. The objective is to estimate the treatment effect for the treated unit, defined as the difference between its observed factual outcome and its unobserved counterfactual outcome. Untreated units help account for unobserved trends in the outcome over time that are unrelated to the policy intervention. The key idea is that an optimally weighted combination of the untreated units, referred to as the synthetic control (SC) unit, serves as an appropriate estimate of the counterfactual outcome. See Figure~\ref{fig:illustration} for an illustration.

SCMs have been widely applied in various empirical studies, such as evaluating the impact of terrorism on GDP \citep{AbadieGardeazabal2003} and the decriminalization of indoor prostitution \citep{Cunningham2017}. For instance, \citet{AbadieDiamondHainmueller2010} applies SCM to study the treatment effect of a comprehensive tobacco control program implemented in California in 1988. While smoking rates in California declined following the program’s implementation, it remained unclear whether this decrease was attributable to the policy. For estimating the treatment effect, annual per-capita cigarette sales data from states that did not implement a similar tobacco control program were used as untreated units. SCM was then applied to construct California’s counterfactual outcome, representing the smoking rate in the absence of the tobacco control program.

A key challenge in SCMs is the assumption, made in most existing studies \citep{Abadie2002}, that the expected outcome of the treated unit can be represented as a weighted sum of the outcomes of untreated units. \citet{Ferman2021} highlights that this assumption leads to a non-negligible bias in the estimator. According to \citet{Ferman2021}, this bias arises from endogeneity in the linear regression model implied by the SCM assumption due to the correlation between the outcomes of untreated units and the error term. This issue is analogous to the endogeneity caused by measurement error \citep{Greene2003Econometric}. We refer to this problem as \emph{implicit endogeneity}.

To address this issue, we propose an alternative approach based on a mixture model assumption rather than a linear relationship among expected outcomes. Specifically, we assume that the distribution of the treated unit’s outcome can be approximated by a weighted combination of the distributions of the untreated units. This assumption further implies that each moment of the treated unit’s outcome can be represented as a weighted sum of the moments of the untreated units. Leveraging this insight, we estimate SC weights by matching the moments of the treated unit’s outcome to those of the untreated units. We refer to our method as the Moment Matching SCM (MMSCM). Unlike existing methods, our proposed estimator is asymptotically unbiased if the mixture model assumption is correct. We validate the effectiveness of our approach through both simulation studies and empirical analysis.

The mixture model assumption has appeared in previous and concurrent studies, including \citet{Wan2018}, \citet{Gunsilius2020}, and \citet{Shi2022}. Our proposed method can be viewed as a simplification of the distributional SCMs introduced by \citet{Gunsilius2020}. While \citet{Gunsilius2020} does not explicitly address the implicit endogeneity problem, they note that their results diverge from those obtained under standard SCMs. This discrepancy arises because, under the mixture model assumption, \citet{Gunsilius2020} provides a consistent estimator for the treatment effect, whereas existing SCM methods do not. We establish a connection between distributional SCMs and the asymptotic bias problem and propose a simplified estimation method based on moment matching.

In summary, our contributions are as follows:
\begin{itemize}
    \item Reexamining the assumptions necessary for treatment effect estimation in SCMs.
    \item Proposing a moment-matching approach for asymptotically unbiased SCMs.
    \item Simplifying the implementation of moment matching in SCMs.
\end{itemize}

We formulate the problem in Section~\ref{sec:problem_setting}. Next, in Section~\ref{sec:recap_scm}, we review the basic idea of SCMs. In Section~\ref{sec:imp_end}, we highlight the problem of implicit endogeneity. In Section~\ref{sec:distscm}, we then propose our method under the mixture model assumption. We investigate the asymptotic properties of this method in Section~\ref{sec:conv_analysis} and present a statistical inference method in Section~\ref{sec:inference}. In Section~\ref{sec:discuss}, we discuss related topics and existing studies. Finally, in Section~\ref{sec:exp}, we examine the performance of our estimators through extensive simulations.

\textbf{Related work.}
Following the pioneering work by \citet{AbadieGardeazabal2003}, numerous studies have advanced the use of SCMs in comparative studies \citep{AbadieDiamondHainmueller2010,AbadieDiamondHainmueller2015}. These SCMs, including their variants such as those by \citet{Doudchenko2016} and \citet{Ben2019}, are based on minimizing the pre-treatment sum of squared errors between the outcome of the treated unit and the weighted sum of those of the untreated units, with some regularization and constraints regarding the weights. We refer to such SCMs as Least-Squares SCMs (LS-SCMs).

The bias of the LS-SCMs strongly depends on whether there exist weights that can predict the outcome of the treated unit without error. In other words, a random outcome of the treated unit must be replicable by the sum of outcomes of untreated units. Since such weights yield a perfect fit in the estimation process using observations from the pre-treatment period, this situation is called pretreatment-fit.

\citet{AbadieDiamondHainmueller2010} demonstrates that under a pre-treatment fit, the bias of the LS-SCM estimator converges to zero asymptotically as the length of the pre-treatment periods $T_0$ increases. However, \citet{Ferman2021} finds that if the pre-treatment fit does not hold, then the LS-SCM estimator has a bias and is not consistent. The cause of this bias is the correlation between the outcomes of the untreated units and the error term in the posited linear model, which we refer to as implicit endogeneity.

Several methods have been proposed to address this bias issue. \citet{Ferman2021OntheProp} demonstrates that we can obtain an asymptotically unbiased estimator under the regime where both the number of untreated units ($J \to \infty$) and the length of the pre-intervention periods ($T_0$) grow large. Concurrently with us, \citet{fry2024method} also discusses the endogeneity problem in SCMs as a cause of the asymptotic bias reported by \citet{Ferman2021}.\footnote{Note that his study became public on arXiv after our study became public on arXiv and at the ICML workshop, and he and we conducted our studies completely independently.} While our formulation and his regarding endogeneity are similar, our approach differs significantly. We address the problem through moment matching, similar to the strategy in \citet{Gunsilius2020}, whereas \citet{fry2024method} employs an instrumental variable approach using outcomes of untreated units not used in SCMs as instruments.

In this study, we focus on mixture models among outcome distributions, an assumption also employed in \citet{Wan2018}, \citet{Gunsilius2020}, \citet{Shi2022}, and \citet{nazaret2023misspecification}. Although these studies introduce mixture models with motivations other than the bias problem, we highlight that such models can effectively circumvent bias under imperfect pre-treatment fit.

Here, we briefly review existing work using mixture models. \citet{Wan2018} points out that SCMs essentially rely on mixture models, distinguishing them from traditional panel data analysis. \citet{Gunsilius2020} proposes distribution SCMs (DiSCo), which estimate SC weights by matching the outcome distributions of the treated unit and the weighted sum of the distributions of the untreated units. \citet{Shi2022} introduces fine-grained models to elucidate hidden assumptions in SCMs and explores connections to linear factor models. \citet{nazaret2023misspecification} extends these insights by analyzing model misspecification under a mixture model formulation.

This study also addresses a question raised by \citet{Gunsilius2020}. While \citet{Gunsilius2020} suggests that standard least squares SCMs try to match the first moment of the outcome distribution (and therefore differ from distributional approaches), we stress that least squares approaches are additionally vulnerable to implicit endogeneity, which can generate bias. Our moment-matching approach avoids this problem by matching a broader set of distributional features.

\begin{table}[t]
    \centering
        \caption{Comparison of SCMs. See the following sections for the definitions of variables and methods.}
    \label{tab:my_label}
    \scalebox{0.65}{
    \begin{tabular}{c|c|c|c|c}
         & Method & Assumption & Asymptotic unbiasedness  
 & Remark\\
    \hline
    Ours & MMSCM & $F_{Y^N_{0,t}}(y) = \sum_{j\in\mathcal{J}^U}w^*_j F_{Y^N_{j,t}}(y)\ \ \ \forall y \in \mathbb{R}$ & Unbiased &  \\
    \hline
        \citet{AbadieGardeazabal2003} & Constrained least squares & $\mathbb{E}[Y^N_{0,t}] = \sum_{j\in\mathcal{J}^U}w^*_j \mathbb{E}[Y^N_{j,t}]$ & Biased  & \\
        \citet{Ferman2021OntheProp} & Constrained least squares & $\mathbb{E}[Y^N_{0,t}] = \sum_{j\in\mathcal{J}^U}w^*_j \mathbb{E}[Y^N_{j,t}]$ & Unbiased & $J\to\infty$ \\
        \citet{Gunsilius2020} & DiSCo & $F^{-1}_{Y^N_{0,t}}(q) = \sum_{j\in\mathcal{J}^U}w^*_j F^{-1}_{Y^N_{j,t}}(q)\ \ \ \forall q \in (0, 1)$  & Unbiased & \\
        \citet{fry2024method} & IV method (GMM) & $\mathbb{E}[Y^N_{0,t}] = \sum_{j\in\mathcal{J}^U}w^*_j \mathbb{E}[Y^N_{j,t}]$ & Unbiased & Use IVs \\
    \end{tabular}
}
\end{table}

\section{Problem Setting}
\label{sec:problem_setting}
Suppose there are $J+1$ units, indexed by $j \in \mathcal{J} \coloneqq \{0, 1, 2, \dots, J\}$, and a time series indexed by $t \in \mathcal{T} = \{1, 2, \dots, T\}$, where $3 \leq T \in \mathbb{N}$. Without loss of generality, assume that the first unit ($j = 0$) is the treated unit, i.e., the unit affected by the policy intervention of interest. The set of potential control units is $\mathcal{J}^U \coloneqq \mathcal{J} \backslash \{0\} = \{1, 2, \dots, J\}$, consisting of units not affected by the intervention. We also assume that our data span $T$ periods and that the intervention occurs at $t = T_0$, with $2 \leq T_0 < T$. Observations in period $t \in \mathcal{T}_0 \coloneqq \{1, 2, \dots, T_0\}$ are from before the intervention, and observations in period $t \in \mathcal{T}_1 \coloneqq \mathcal{T} \backslash \mathcal{T}_0$ are from after the intervention. Let $T_1 \coloneqq |\mathcal{T}_1| = T - T_0$.

\subsection{Potential Outcomes}
Following the Neyman-Rubin causal model \citep{Neyman1923,Rubin1974}, we define the potential outcomes of the policy intervention. For each unit $j \in \mathcal{J}$ and period $t \in \mathcal{T}$, there are two outcomes of interest, $\left(Y^I_{j,t}, Y^N_{j,t}\right) \in \mathbb{R}^2$, corresponding to scenarios with and without the intervention. For each $j \in \mathcal{J}$ and $t \in \mathcal{T}$, let $F^I_{j,t}(y)$ and $F^N_{j,t}(y)$ be the cumulative distribution functions (CDFs) for $Y^I_{j,t}$ and $Y^N_{j,t}$, respectively.

\subsection{Observations}
We next define our observations. For each unit $j \in \mathcal{J}$, only one of the two potential outcomes can be observed, corresponding to what actually happened. Specifically, for each $j \in \mathcal{J}$, we observe $Y_{j,t} \in \mathbb{R}$ such that
\begin{align*}
    Y_{0,t} = \begin{cases}
    Y^N_{0,t} & \forall t \in \mathcal{T}_0,\\
    Y^I_{0,t} & \forall t \in \mathcal{T}_1,
    \end{cases}
    \quad \mathrm{and} \quad
    Y_{j,t} = Y^N_{j,t} 
    \quad \forall j \in \mathcal{J}^U,   \forall t \in \mathcal{T}.
\end{align*}
We aim to estimate the effect of the policy intervention by predicting $Y^I_{0,t}$ from $\left(Y^N_{j,t}\right)_{j \in \mathcal{J}^U}$ and comparing that prediction to the observed $Y^N_{0,t}$ for $t \in \mathcal{T}_1$. We define the treatment effect more formally in the next subsection.

\subsection{Treatment Effect}
In SCMs, there are several ways to define the treatment effect $\tau_{0,t} \in \mathbb{R}$ for each $t \in \mathcal{T}_1$ (i.e., after the intervention). First, for each $t \in \mathcal{T}_1$, we can define the treatment effect for the treated unit $j = 0$ as
\begin{align*}
    \tau_{0,t} \coloneqq \mathbb{E}[Y^I_{0,t}] - \mathbb{E}[Y^N_{0,t}].
\end{align*}

Another definition is based on the following linear factor model:
\begin{definition}
\label{def:linear_contextual}
We say that $(Y^I_{j,t}, Y^N_{j,t})$ follows the linear factor model if:
\begin{align}
\label{eq:latent}
    &Y^N_{j, t} = c_j + \delta_t + \lambda_t \mu_j + \epsilon_{j, t},\quad \mathbb{E}_{j, t}[\epsilon_{j, t}] = 0,\quad \forall t \in \mathcal{T},\quad \forall j \in \mathcal{J},\\
    &Y^I_{j, t} = \tau_{j, t} + Y^N_{j, t},\nonumber
\end{align}
where $\delta_t$ is an unobserved common factor with constant factor loadings across units, $c_j$ is an unknown time-invariant fixed effect, $\lambda_t$ is a $(1 \times F)$ vector of unobserved common factors, $\mu_j$ is a $(F \times 1)$ vector of unknown factor loadings, and $\epsilon_{j, t}$ is an unobserved idiosyncratic shock. Here, $c_j$, $\mu_j$, and $\tau_{0,t}$ are non-random variables, while $\delta_t$, $\lambda_t$, and $\epsilon_{j,t}$ are random variables.
\end{definition}
Under the linear factor model, the treatment effect is given by $\tau_{0,t} = Y^I_{0,t} - Y^N_{0,t}$, a non-random variable by definition. Consequently, the expression $\tau_{0,t} \coloneqq \mathbb{E}[Y^I_{0,t}] - \mathbb{E}[Y^N_{0,t}]$ also holds under the factor model.

Comparative case studies aim to evaluate $\tau_{0,t}$ by predicting $Y^N_{0,t}$ for $t > T_0$, the counterfactual outcome that would have been observed for the treated unit had the intervention not occurred. Because the intervention takes place after $t = T_0$, $Y^N_{0,t}$ is a counterfactual outcome. In settings where data consist of only a few units (e.g., regions or countries), it can be difficult to find a single untreated unit that provides a suitable comparison. For this problem, SCMs enable us to construct a counterfactual outcome using a predictor based on the outcomes of the untreated units. Under suitable model assumptions that relate $Y^N_{0,t}$ to $\left(Y^N_{j,t}\right)_{j \in \mathcal{J}^U}$, we can construct the predictor using observations from the pre-intervention period and then predict the unobserved counterfactual outcome in the post-intervention period. Consequently, we estimate the treatment effect by comparing the predicted $Y^N_{0,t}$ with the observed $Y^I_{0,t}$.

\textbf{Distributional Treatment Effects.} In program evaluation, we are often interested not only in the scalar treatment effect but also in the distributions of outcomes \citep{Maier2011} or their functionals, such as quantile treatment effects (QTEs) and expected social welfare \citep{Abadie2002}. By applying our method, we can also estimate the distribution of the counterfactual outcome, $p^N_{0,t}(y)$, for $t \in \mathcal{T}_1$. 
The treatment effect is then captured by the difference between $p^I_{0,t}(y)$ (the distribution of the actual outcome) and $p^N_{0,t}(y)$, or by functionals of these distributions, such as QTEs and expected social welfare. Depending on the application, we may also adjust the target density, for example, by averaging across all post-intervention periods: $\overline{p}^N_{0}(y) = \frac{1}{T_1}\sum_{t\in\mathcal{T}_1}p^N_{0,t}(y)$. We refer to the treatment effects related to these distributions as \emph{distributional treatment effects} \citep{ParkShalit2021, KallusOprescu2022, chikahara2022feature}. In the context of SCMs, \citet{Gunsilius2020} and \citet{Chen2020} have proposed methods for estimating such distributional treatment effects.

\section{Recap of SCMs}
\label{sec:recap_scm}
This section reviews the original SCM proposed by \citet{AbadieGardeazabal2003}. 
To estimate the treatment effect $\tau_{0,t}$, we predict the counterfactual outcome $Y^N_{0,t}$ using a weighted sum of $\{Y^N_{j,t}\}_{j=1}^J$ for $t\in\mathcal{T}_1$. Specifically, SCMs predict $Y^N_{0,t}$ as
\[
\widehat{Y}^N_{0,t} \coloneqq w_0 + \sum_{j\in\mathcal{J}^U} w_j Y^N_{j,t},
\]
where $w_0 \in\mathbb{R}$ and $w_j \in\mathbb{R}$ $(j\in\mathcal{J}^U)$ are the SC weights.

Several methods exist for estimating the SC weights. In the SCM proposed by \citet{AbadieGardeazabal2003}, these weights are determined by minimizing the squared error between $Y^N_{0,t}$ and the weighted sum of $Y^N_{1,t},\dots,Y^N_{J,t}$. Define
\[
\Delta^J \coloneqq \left\{\bm{w} = (w_1\ w_2\ \dots\ w_J)^\top \in [0, 1]^J \mid \sum_{j\in\mathcal{J}^U}w_j = 1\right\}.
\]
Then, \citet{AbadieGardeazabal2003} estimate the SC weights as
\[
\bm{w}^{\mathrm{Abadie}} = \argmin_{\bm{w}\in\Delta^J} \sum_{t\in\mathcal{T}_0}\left(Y_{0, t} - \sum_{j\in\mathcal{J}^U} w_jY^N_{j, t}\right)^2.
\]
Recall that we referred to SCMs using this approach as the Least-Squares-SCMs (LS-SCMs). 
The counterfactual outcome is then predicted as
\[
\widehat{Y}^{N, \mathrm{Abadie}}_{0, t} = \widehat{w}^{\mathrm{Abadie}}_0 + \sum_{j\in\mathcal{J}^U}\widehat{w}^{\mathrm{Abadie}}_jY^N_{j,t}.
\]
Using $\widehat{Y}^{N, \mathrm{Abadie}}_{0, t}$, we estimate the treatment effect as
\[
\widehat{\tau}^{\mathrm{Abadie}}_{t} = Y^I_{0, t} - \widehat{Y}^{N, \mathrm{Abadie}}_{0, t}.
\]

The literature on inference for SCMs has grown significantly, extending the original proposals by \citet{AbadieGardeazabal2003} and \citet{AbadieDiamondHainmueller2010}. Notable contributions include works by \citet{Kathleen2020}, \citet{Shaikh2021}, \citet{CattaneoFeng2021}, and \citet{Chernozhukov2021}. \citet{Doudchenko2016} proposes estimating the SC weight as
\[
\widetilde{\bm{w}}^{\mathrm{DI}} = \argmin_{\widetilde{\bm{w}} = (w_0\ w_1\ \cdots\ w_J)^\top \in \mathbb{R}^{J+1}} \sum_{t\in\mathcal{T}_0}\left(Y_{0, t} - w_0 - \sum_{j\in\mathcal{J}^U} w_jY_{j, t}\right)^2 + \lambda_1 \|\widetilde{\bm{w}}\|_1 + \lambda_2\|\widetilde{\bm{w}}\|_2,
\]
where $\lambda_1 > 0$ and $\lambda_2 > 0$ are regularization parameters. 

\section{Identification and Implicit Endogeneity}
\label{sec:imp_end}
This section addresses the identification problem of treatment effects and points out the problem of implicit endogeneity in LS-SCMs.

\subsection{Linear Models in Expected Outcomes}
We revisit the assumptions underlying SCMs. In many existing studies, a linear relationship is assumed between $Y^N_{0,t}$ and $Y^N_{1,t},\dots, Y^N_{J,t}$. One possibility is positing the existence of $\bm{w}^* \in \Delta^J$ \citep{AbadieDiamondHainmueller2010} such that
\[
Y^N_{0,t} = \sum_{j\in[J]} w^*_j Y^N_{j, t}.
\] 
This assumption, often called the perfect pre-treatment fit \citep{Ferman2021}, implies that the stochastic process $\left(Y^N_{0,t}\right)_{t\in \mathcal{T}_0}$ can be fully replicated by $\sum_{j\in[J]} w^*_j Y^N_{j, t}$ for $t\in \mathcal{T}_0$. In linear factor models where $Y^N_{j, t} = c_j + \delta_t + \lambda_t \mu_j + \epsilon_{j, t}$, this assumption might imply $\epsilon_{0, t} = \sum_{j\in[J]} w^*_j \epsilon_{j, t}$, which is often considered unrealistic. Note that this is a specific case and not necessary for $Y^N_{0,t} = \sum_{j\in[J]} w^*_j Y^N_{j, t}$. 

A more natural assumption is a linear model between $\mathbb{E}_{0, t}[Y^N_{0, t}]$ and $(\mathbb{E}_{j, t}[Y^N_{j, t}])_{j\in\mathcal{J}^U}$.

\begin{assumption}
\label{asm:linearity}
There exists a set $\widetilde{\Phi}\subset \Delta^J$ such that for all $\bm{w}^* = (w^*_j)_{j\in\mathcal{J}^U} \in \widetilde{\Phi}$,
\begin{align}
\label{eq:linear_exp}
\mathbb{E}_{0, t}[Y^N_{0, t}] = \sum_{j\in\mathcal{J}^U} w^*_j \mathbb{E}_{j, t}[Y^N_{j, t}].
\end{align}
\end{assumption}

Compared to Assumption~\ref{asm:linearity}, the assumption $Y_{0, t} = \sum_{j\in\mathcal{J}^U} w^*_j Y_{j, t}$ is stronger and less plausible, given that $Y_{j,t}$ is random. 

\subsection{Asymptotic Bias in Linear Factor Models}
\label{sec:ferman}
We now consider the linear factor model introduced in Definition~\ref{def:linear_contextual}, along with Assumption~\ref{asm:linearity}. This is a specific case of \eqref{eq:linear_exp}. Under the linear factor model, we adopt the following more specialized assumption:

\begin{assumption}
\label{asm:linearity2}
Under the linear factor model from Definition~\ref{def:linear_contextual}, there exists a set $\widetilde{\Phi}\subset \Delta^J$ such that for all $\bm{w}^* = (w^*_j)_{j\in\mathcal{J}^U} \in \widetilde{\Phi}$,
\begin{align*}
c_0 = \sum_{j\in[J]}w^*_j c_j 
\quad \text{and} \quad 
\mu_0 =  \sum_{j\in[J]}w^*_j \mu_j 
\quad \forall t \in \mathcal{T},\ \forall j \in\mathcal{J}.
\end{align*}    
\end{assumption}

Assumption~\ref{asm:linearity2} is a specific case of Assumption~\ref{asm:linearity}; if Assumption~\ref{asm:linearity} holds, then Assumption~\ref{asm:linearity2} also holds under the linear factor model in Definition~\ref{def:linear_contextual}. 

Under these assumptions, \citet{Ferman2021} shows that the SC estimator $\widehat{\tau}^{\mathrm{Abadie}}_{0,t}$ suffers from bias and is not consistent for $\tau_{0, t}$ when $c_j$ is non-zero. We also adopt the following assumption regarding the linear factor model.

\begin{assumption}[Common and idiosyncratic shocks]
\label{asm:common_idio} 
For each $j\in\mathcal{J}$, the linear factor model in Definition~\ref{def:linear_contextual} satisfies the following as $T_0 \to \infty$: 
\begin{itemize}
    \item $\frac{1}{T_0}\sum_{t\in\mathcal{T}_0} \lambda_t \xrightarrow{\mathrm{p}}0$.
    \item $\frac{1}{T_0}\sum_{t\in\mathcal{T}_0} \epsilon_{j, t} \xrightarrow{\mathrm{p}}0$.
    \item $\frac{1}{T_0}\sum_{t\in\mathcal{T}_0} \lambda^\top_t \lambda_t \xrightarrow{\mathrm{p}} \Omega_0$, where $\Omega_0$ is positive semidefinite.
    \item $\frac{1}{T_0}\sum_{t\in\mathcal{T}_0} \epsilon_{j, t} \epsilon^\top_{j, t} \xrightarrow{\mathrm{p}} \sigma^2_{j, \epsilon} I_{J+1}$, where $\sigma^2_{j, \epsilon} \geq 0$ is constant.
    \item $\frac{1}{T_0}\sum_{t\in\mathcal{T}_0} \epsilon_{j, t} \lambda_t \xrightarrow{\mathrm{p}}0$.
\end{itemize}
\end{assumption}

Under Assumptions~\ref{asm:linearity2} and \ref{asm:common_idio}, \citet{Ferman2021} concludes that the LS-SCM proposed by \citet{AbadieGardeazabal2003} does not yield a consistent estimator for $\tau_{0, t}$, as the following proposition. 

\begin{proposition}[Proposition~1 in \citet{Ferman2021}]
Under Assumptions~\ref{asm:common_idio}--\ref{asm:linearity2}, we have
\[
\bm{w}^{\mathrm{Abadie}} \xrightarrow{\mathrm{p}} \overline{\bm{w}}^{\mathrm{Abadie}} \coloneqq  \left(\overline{w}^{\mathrm{Abadie}}_j\right)_{j\in\mathcal{J}^U}
\]
as $T_0 \to \infty$, where 
\[
(c_0, \mu_0) \neq \left(\sum_{j\in\mathcal{J}^U}c_j \overline{w}^{\mathrm{Abadie}}_j, \sum_{j\in\mathcal{J}^U}\mu_j \overline{w}^{\mathrm{Abadie}}_j\right),
\]
unless $\sigma^2_{\epsilon} = 0$ or $\widetilde{\Phi} \cap \argmin_{\bm{w}\in\Delta^J} \{\bm{w}^\top \bm{w}\} \neq \emptyset$. Furthermore, for $t\in\mathcal{T}_1$,
\[
\widehat{\tau}^{\mathrm{Abadie}}_{0, t} \xrightarrow{\mathrm{p}} \tau_{0, t} 
+ \lambda_t\left(\mu_{0} - \sum_{j\in\mathcal{J}^U}\mu_j \overline{w}^{\mathrm{Abadie}}_j\right)
+ \left(c_0 - \sum_{j\in\mathcal{J}^U}c_j \overline{w}^{\mathrm{Abadie}}_j\right)
+ \left(\epsilon_{0, t} - \sum_{j\in\mathcal{J}^U}\epsilon_{j, t} \overline{w}^{\mathrm{Abadie}}_j\right)
\]
as $T_0 \to \infty$.
\end{proposition}

To alleviate this bias problem, \citet{Ferman2021} proposes the demeaned SC estimator:
\[
    \widehat{\bm{w}}^{\mathrm{FP}} = \argmin_{\widehat{\bm{w}}^{\mathrm{FP}}\in\Delta^J}\frac{1}{T_0}\sum_{t\in\mathcal{T}_0}\left\{\left(Y^N_{0, t} - \overline{Y}^N_0\right) - \sum_{j\in\mathcal{J}^U}w_j\left(Y^N_{j, t} - \overline{Y}^N_j\right)\right\}^2,
\]
where $\overline{Y}^N_j = \frac{1}{T_0}\sum_{t\in\mathcal{T}_0}Y_{j, t}$. They then estimate $\tau_{0, t}$ as
\[
\widehat{\tau}^{\mathrm{FP}}_{0, t} 
= Y^I_{0, t} - \overline{Y}^N_0 
- \sum_{j\in\mathcal{J}^U}\widehat{w}^{\mathrm{FP}}_j\left(Y^N_{j, t} - \overline{Y}^N_j\right).
\]
They show that this approach can reduce the bias seen in the classical LS-SCM estimator. However, there still remains bias in the estimator, which does not vanish even if $T_0 \to \infty$. 

\subsection{Implicit Endogeneity}
We now interpret the asymptotic bias found by \citet{Ferman2021} as an endogeneity problem. We consider \eqref{eq:linear_exp}, that is, $\mathbb{E}_{0, t}[Y^N_{0, t}] = \sum_{j\in\mathcal{J}^U} w^*_j \mathbb{E}_{j, t}[Y^N_{j, t}]$, which is more general than the linear factor model.

To illustrate, we consider a simpler case with the least-squares estimator rather than $\bm{w}^{\mathrm{Abadie}}$:
\[
\widehat{\bm{w}}^{\mathrm{LS}} = \argmin_{\bm{w}\in\mathbb{R}^J} \frac{1}{T_0}\sum_{t\in\mathcal{T}_0}\left(Y_{0, t} - \sum_{j\in\mathcal{J}^U} w_j Y_{j, t}\right)^2.
\]
We show that $\widehat{\bm{w}}^{\mathrm{LS}}$ is asymptotically biased due to endogeneity. Define $\epsilon_{j, t} = Y_{j, t} - \mathbb{E}_{j, t}[Y_{j, t}]$. Under \eqref{eq:linear_exp}, we obtain
\begin{align*}
Y_{0, t} = \sum_{j\in\mathcal{J}^U} w^*_j Y_{j, t} + \nu_{t}, \quad \text{where} \quad 
\nu_{t} = - \sum_{j\in\mathcal{J}^U} w^*_j \epsilon_{j, t} + \epsilon_{0, t}.
\end{align*}
Noting that $\mathbb{E}[\nu_{t} Y_{j, t}] \neq 0$, we see that there is endogeneity, as the regressor and the error term are correlated. Endogeneity is known to induce bias in least-squares estimation. Let $Q^*$ be a $(J \times J)$ matrix whose $(i, j)$-element is $\frac{1}{T_0}\sum_{t\in\mathcal{T}_0}\mathbb{E}_{i, t}[Y_{i, t}]\mathbb{E}_{j, t}[Y_{j, t}]$, and let $\Sigma$ be a $J \times J$ diagonal matrix whose $(j,j)$-element is $\frac{1}{T_0}\sum_{t\in\mathcal{T}_0} \sigma^2_{j, t}$.

\begin{theorem} 
As $T_0\to \infty$, $\widehat{\bm{w}}^{\mathrm{LS}} \xrightarrow{\mathrm{p}} \overline{\bm{w}}^{\mathrm{LS}}$, where 
\[
    \overline{\bm{w}}^{\mathrm{LS}} 
    = \bm{w}^* - \left(Q^* + \Sigma\right)^{-1}\Sigma \bm{w}^*.
\]
\end{theorem}

This result follows from Section~5.6 of \citet{Greene2003Econometric} and implies that, under \eqref{eq:linear_exp}, there is an asymptotic bias in $\widehat{\bm{w}}^{\mathrm{LS}}$; in other words, least-squares-type estimators do not converge to the true $\bm{w}^*$. 

Thus, the identification of treatment effects faces a challenge due to implicit endogeneity. While the estimator $\widehat{\bm{w}}^{\mathrm{Abadie}}_n$ remains standard in many applications, it inherits this endogeneity issue. Such concerns have been noted in prior works \citep{Chernozhukov2021,Ferman2021,Ferman2021OntheProp}, and \citet{fry2024method} independently frames the bias as an endogeneity problem, proposing an instrumental variable method to address it.


\textbf{Bias is caused by a measurement error.} As we showed above, the source of the implicit endogeneity is the measurement error. Since we assume $\mathbb{E}_{0, t}[Y^N_{0, t}] = \sum_{j\in\mathcal{J}^U} w^*_j \mathbb{E}_{j, t}[Y^N_{j, t}]$ as a regression model, the regressors are $(\mathbb{E}_{j, t}[Y^N_{j, t}])_{j \in \mathcal{J}^U}$. However, in practice, we can only use $(Y_{j, t})$ as the regressors. Because the regressors contain measurement errors $(\mathbb{E}_{j, t}[Y^N_{j, t}] - Y_{j, t})_{j \in \mathcal{J}^U}$, endogeneity arises, as is commonly discussed in elementary textbooks \citep{Greene2003Econometric}.

\color{black}

\section{SCM by Moment Matching}
\label{sec:distscm}
As discussed in Section~\ref{sec:imp_end}, Assumption~\ref{asm:linearity} can be weak or incompatible for identifying $\bm{w}^*$ when using least squares. We, therefore, adopt a different assumption for identification. Specifically, we propose that the distribution of $Y^N_{0, t}$ can be replicated by a weighted combination of the distributions of $Y^N_{j, t}$, and we introduce an SCM based on moment matching.

\subsection{Identification and Estimation Strategy}
Section~\ref{sec:imp_end} and the results in \citet{Ferman2021} indicate that combining 
\[
\mathbb{E}_{0, t}\left[Y_{0, t}\right] 
= \sum_{j\in\mathcal{J}^U} w^*_j \mathbb{E}_{j, t}\left[Y_{j, t}\right]
\]
with least squares estimators do not yield consistent estimators of $\bm{w}^*$. We thus adopt mixture models as a more theoretically favorable but still realistic assumption.

\textbf{Identification.}
We begin by positing that the distribution of the target unit's outcome $Y^N_{0, t}$ can be expressed as a linear combination of the distributions of $Y^N_{1, t}, \dots, Y^N_{J, t}$. For simplicity, in this subsection, we assume that $Y^N_{j,t}$ is a continuous random variable with a probability density function (PDF). Under this scenario, we suppose that the density of $Y^N_{0, t}$ can be replicated by a weighted sum of the densities of $Y^N_{j, t}$:
\begin{align}
\label{eq:mixture_model}
    p^N_{0,t}(y) = \sum_{j\in\mathcal{J}^U} w^*_j p^N_{j,t}(y) \qquad \forall t \in \mathcal{T}.
\end{align}

\textbf{Estimation.}
We next discuss how to estimate the SC weights that satisfy \eqref{eq:mixture_model}. One approach employs optimal transport to minimize the Wasserstein distance \citep{Gunsilius2020}. In this study, we propose a simpler estimation procedure, as the approach of \citet{Gunsilius2020} is more complex than that of the original SCMs.

There are various ways to estimate $(w^*_j)_{j\in\mathcal{J}^U}$. We consider the following:

\begin{itemize}
    \item Distribution divergence (distance) minimization.
    \item Characteristic function matching.
    \item Moment matching.
\end{itemize}

A straightforward approach to estimating $(w^*_j)_{j\in\mathcal{J}^U}$ is to minimize the divergence or distance between two distributions whose densities are given by $p^N_{0,t}(y)$ and $\sum_{j\in\mathcal{J}^U} w_j p^N_{j,t}(y)$, given $(w_j)_{j\in\mathcal{J}^U}$. For the divergence or distance measures, we may use, for example, the Kullback-Leibler (KL) divergence or the Wasserstein distance. The KL divergence can be estimated nonparametrically using density ratio techniques (assuming the densities exist) \citep{Sugiyama:2012:DRE:2181148,Kato2021bregman}. Alternatively, we may estimate the weights by minimizing the maximum mean discrepancy.

The second approach is to match characteristic functions. 
Since the mixture model \eqref{eq:mixture_model} holds, we have
$
\mathbb{E}_{0,t}\left[\exp\left(\sqrt{-1} \xi Y^N_{0,t}\right)\right] 
= \sum_{j\in\mathcal{J}^U} w^*_j  \mathbb{E}_{j,t}\left[\exp\left(\sqrt{-1} \xi Y^N_{j,t}\right)\right]
$ for all $t \in \mathcal{T}$ and $\xi\in\mathbb{R}$.  
where $\sqrt{-1}$ denotes the imaginary unit. The LHS is the characteristic function (CHF) of $p^N_{0,t}(y)$, and the RHS is that of $\sum_{j\in\mathcal{J}^U} w_j p^N_{j,t}(y)$, given $(w_j)_{j\in\mathcal{J}^U}$. Since CHFs uniquely determine the corresponding distributions, we can estimate $(w^*_j)_{j\in\mathcal{J}^U}$ by matching the CHFs. Such matching can be implemented using the method of moments, for instance, the approach proposed in \citet{Carrasco2017efficientestimation}.

The last approach is to match the moments of the distributions $p^N_{0,t}(y)$ and $\sum_{j\in\mathcal{J}^U} w_j p^N_{j,t}(y)$. Intuitively, if we find $(w_j)_{j\in\mathcal{J}^U}$ such that all the moments of the distributions match, we can interpret it as an estimate of $(w^*_j)_{j\in\mathcal{J}^U}$. This idea can also be motivated from the perspective of CHF matching, as noted in Remark~\ref{rem:chfmoment}. Note that to justify this approach, we must ensure that the moments uniquely determine the distributions. This is the classical moment problem \citep{Stchmudgen2020lecturesmomentproblem}, which has been extensively studied. It is known that certain conditions are required to guarantee uniqueness.

Among the three approaches, we focus on the last one, moment matching. As we show later, it is sufficient to estimate $(w^*_j)_{j\in\mathcal{J}^U}$ by matching sample means of polynomials of $Y_{a,j}$. Compared to the other approaches, the implementation of moment matching is significantly simpler. Note that it cannot be applied to all general distributions without additional assumptions since the moments may not be uniquely determined for each distribution. However, if we assume boundedness of $Y_{a,j}$, we can ensure that the distributions are uniquely determined by their moments \citep{Hausdorff1921summationsmethodenund}. In this study, for simplicity, we adopt this moment matching approach under the assumption of bounded random variables.
\begin{remark}[Distribution SCMs using divergence or distance minimization.]
Since we are interested in finding $(w_j)_{j\in\mathcal{J}^U}$ such that two distributions match, it is sufficient to use a measure that guarantees the distributions are identical when the divergence or distance is zero. That is, we do not need to consider symmetry or deficient support, which often motivate the use of the Wasserstein distance instead of the KL divergence. Therefore, while \citet{Gunsilius2020} introduces distributional SCM using the Wasserstein distance, we believe that there is no strong advantage to using it over other divergence or distance measures.
\end{remark}

\begin{remark}[From CHF matching to moment matching]
\label{rem:chfmoment}
We can estimate $\bm{w}^*$ by bounding the difference between the CHFs of $Y^N_{0,t}$ and the weighted sum of $Y^N_{j,t}$. Specifically, we have
\begin{align*}
    &\int_{\xi \in (-h, h)} \left|\mathbb{E}_{0,t}\left[\exp(\sqrt{-1} \xi Y^N_{0,t})\right] - \sum_{j\in\mathcal{J}^U} w_j \mathbb{E}_{j,t}\left[\exp(\sqrt{-1} \xi Y^N_{j,t})\right]\right| \mathrm{d}\xi\\
    &= \int_{\xi \in (-h, h)} \Biggl|\sum_{\gamma\in\{0,1,2,\dots, G\}} 
    \frac{\sqrt{-1}^\gamma \xi^\gamma}{\gamma !} \left(\mathbb{E}_{0,t}\left[\left(Y^N_{0,t}\right)^\gamma \right] - \sum_{j\in\mathcal{J}^U} w_j \mathbb{E}_{j,t}\left[\left(Y^N_{j,t}\right)^\gamma\right]\right) + \epsilon  |\xi^G|\Biggr|\mathrm{d}\xi\\
    &\leq \int_{\xi \in (-h, h)} \Biggl(\sum_{\gamma\in\{1,2,\dots, G\}}
    \frac{|\sqrt{-1}^\gamma| |\xi|^\gamma}{\gamma !} 
    \left|\mathbb{E}_{0,t}\left[\left(Y^N_{0,t}\right)^\gamma \right] 
    - \sum_{j\in\mathcal{J}^U} w_j \mathbb{E}_{j,t}\left[\left(Y^N_{j,t}\right)^\gamma\right]\right|
    + \epsilon |\xi^G|\Biggr) \mathrm{d}\xi\\
    &= \sum_{\gamma\in\{1,2,\dots, G\}} \frac{2 h^{\gamma + 1}}{(\gamma + 1)!}
    \left|\mathbb{E}_{0,t}\left[\left(Y^N_{0,t}\right)^\gamma \right] 
    - \sum_{j\in\mathcal{J}^U} w_j \mathbb{E}_{j,t}\left[\left(Y^N_{j,t}\right)^\gamma\right]\right|
    +  \frac{2 \epsilon}{G+1} h^{G+1}.
\end{align*}
Minimizing
\[
\sum_{\gamma\in\{1,2,\dots, G\}}
\frac{2 h^{\gamma + 1}}{(\gamma + 1)!}
\left|\frac{1}{T_0}\sum_{t\in\mathcal{T}_0}\left(\left(Y^N_{0,t}\right)^\gamma  
- \sum_{j\in\mathcal{J}^U} w_j \left(Y^N_{j,t}\right)^\gamma\right)\right| 
+ \frac{2 \epsilon}{G+1} h^{G+1}
\]
thus yields an accurate estimate of $\bm{w}^*$ that ensures
\[
\int_{\xi \in (-h, h)} \left|\mathbb{E}_{0,t}\left[\exp(\sqrt{-1} \xi Y^N_{0,t})\right] 
- \sum_{j\in\mathcal{J}^U} w_j \mathbb{E}_{j,t}\left[\exp(\sqrt{-1} \xi Y^N_{j,t})\right]\right| \mathrm{d}\xi = 0.
\]
This approach can be viewed as a variant of the method of moments, where we match all moments of $Y^N_{0,t}$ with those of the weighted sum of the untreated units.
\end{remark}

\begin{remark}[Reports in \citet{Gunsilius2020}]
Distributional SCMs are also considered in \citet{Gunsilius2020}. However, the author does not emphasize the fact that distributional SCMs can estimate the treatment effect in an asymptotically unbiased manner, whereas original SCMs such as \citet{AbadieGardeazabal2003} yield biased estimators. For example, \citet{Gunsilius2020} states that ``the distributional synthetic controls method finds optimal weights that replicate all moments of the target distribution as closely as possible. In contrast, applying the classical method to averages of the distributions will find the optimal weights based on replicating the first moment.'' This claim may be accurate, as the classical method cannot yield unbiased and consistent estimates of the optimal weights for replicating the first moment unless, for example, we assume that the outcomes are nonrandom variables.
\end{remark}

\color{black}

\subsection{Core Assumptions}
We now state our core assumption for identifying the SC weights.

\begin{assumption}
\label{asm:bounded}
    The outcomes $(Y^N_{j, t})_{j\in\mathcal{J}, t\in\mathcal{T}}$ are bounded \mkato{as $Y^N_{j, t}\in [0, 1]$}.
\end{assumption}

\begin{assumption}
\label{asm:core_asum}
There exists $\widetilde{\Phi}^* \subset \Delta^J$ such that for all $\bm{w}^* \in \widetilde{\Phi}^*$,
\[
        F_{Y^N_{0,t}}(y) = \sum_{j\in\mathcal{J}^U}w^*_j  F_{Y^N_{j,t}}(y)\quad \forall y \in \mathbb{R}.
\]
\end{assumption}

Assumption~\ref{asm:bounded} ensures that $Y^N_{j, t}$ has a finite $G$-th moment for any $G\in\mathbb{N}$. If $Y^N_{j, t}$ is bounded but not in $[0, 1]$, we can normalize $Y^N_{j, t}$ to satisfy this assumption without loss of generality. Assumption~\ref{asm:core_asum} has been employed in earlier work, including \citet{Wan2018}, \citet{Gunsilius2020}, and \citet{Shi2022}. \citet{Wan2018} notes that SCMs effectively rely on Assumption~\ref{asm:core_asum}. \citet{Gunsilius2020} similarly assumes \eqref{eq:mixture_model} and estimates $\bm{w}^*$ by minimizing the Wasserstein distance, while \citet{Shi2022} introduces fine-grained models that, combined with Assumption~\ref{asm:core_asum}, can justify standard SCMs. We provide further discussion of these fine-grained models in Section~\ref{sec:just_mm}.

Although mixture models and linear factor models differ, they are closely related \citep{Shi2022,nazaret2023misspecification}. In the fine-grained models of \citet{Shi2022}, certain implicit assumptions underlie linear factor models, suggesting that mixture model assumptions can also be valid for these settings. Independently, \citet{nazaret2023misspecification} arrive at a similar conclusion. We detail these connections in Section~\ref{sec:just_mm}.

\subsection{Estimation of SC Weights}
Under the mixture model assumption, we propose estimating SC weights by matching moment functions. We refer to our approach as the Momemnt Matching SCM (MMSCM).

Let $v_\gamma \in (0, \infty)$ for $\gamma = 1, 2, \dots, G$ be a fixed weight sequence. We then estimate the SC weights as
\begin{align}
\label{eq:gmm}
\widehat{\bm{w}}^{\mathrm{MMSCM}}_{T_0, G} \coloneqq  \left(\widehat{w}^{\mathrm{MMSCM}}_{j, T_0, G}\right)_{j\in\mathcal{J}^U} 
\coloneqq \argmin_{\bm{w}\in \Delta^J} 
   \sum_{\gamma=1}^G
   v_\gamma 
   \left|\frac{1}{T_0}\sum_{t\in\mathcal{T}_0}\left(\left(Y^N_{0,t}\right)^\gamma  
   - \sum_{j\in\mathcal{J}^U} w_j \left(Y^N_{j,t}\right)^\gamma\right)\right|.
\end{align}

We may use any weights $v_\gamma \in (0, \infty)$ for $\gamma = 1, 2, \dots, G$. For example, based on Remark~\ref{rem:chfmoment}, we can choose $v_\gamma = \frac{2 h^{\gamma + 1}}{(\gamma + 1)!}$. In this case, we estimate $(w^*_j)_{j\in\mathcal{J}^U}$ as
\begin{align*}
\widehat{\bm{w}}^{\mathrm{MMSCM}}_{T_0, G} \coloneqq  \left(\widehat{w}^{\mathrm{MMSCM}}_{j, T_0, G}\right)_{j\in\mathcal{J}^U} 
\coloneqq \argmin_{\bm{w}\in \Delta^J} 
   \sum_{\gamma=1}^G
   \frac{2 h^{\gamma + 1}}{(\gamma + 1)!} 
   \left|\frac{1}{T_0}\sum_{t\in\mathcal{T}_0}\left(\left(Y^N_{0,t}\right)^\gamma  
   - \sum_{j\in\mathcal{J}^U} w_j \left(Y^N_{j,t}\right)^\gamma\right)\right|.
\end{align*}

Using $\widehat{\bm{w}}^{\mathrm{MMSCM}}_{T_0, G}$, we can then estimate both the treatment effect and the distributional treatment effect, as discussed in the following subsections.

\color{black}

\subsection{Treatment Effect Estimation}
Once $\widehat{\bm{w}}^{\mathrm{MMSCM}}_{T_0, G}$ is obtained, we estimate the treatment effect similarly to standard SCMs:
\begin{align}
\label{eq:att_est}
\widehat{\tau}^{\mathrm{MMSCM}}_{0, t} \coloneqq  Y^I_{0, t} 
  - \sum_{j\in\mathcal{J}^U}\widehat{w}^{\mathrm{MMSCM}}_{j, T_0, G} Y_{j, t}.
\end{align}
We call this estimator the Moment Matching SCM (MMSCM) estimator. Algorithm~\ref{alg:main} summarizes our proposed estimator.


\begin{remark}[Multiple treatments]
We can extend our method to the case with multiple treatments $K \geq 2$. Assume that there are units $- \widetilde{J}, \cdots, -1$ that receive treatments $- \widetilde{J}, \cdots, -1$, while unit $0$ receives treatment $1$. In such cases, how we assume the DGP of the outcomes is crucial. For example, if we assume that the outcomes of the treated units follow the same distribution, it is sufficient to take the average over the treated group; that is, we solve
$\min_{\bm{w}\in \Delta^J} 
   \sum_{\gamma=1}^G
  v_\gamma
   \left|\frac{1}{T_0}\sum_{t\in\mathcal{T}_0}\left(\frac{1}{\widetilde{J} + 1}\sum_{\widetilde{j} \in \{0, -1, \dots, \widetilde{J}\} }\left(Y^N_{\widetilde{j}, t}\right)^\gamma  
   - \sum_{j\in\mathcal{J}^U} w_j \left(Y^N_{j,t}\right)^\gamma\right)\right|$.
We can also consider other situations in this setting, such as differential treatment timing. These extensions remain an open issue in the general context of SCM, and are not limited to the method proposed in this study.
\end{remark}

\color{black}

\begin{algorithm}[t]
   \caption{MMSCM.}
   \label{alg:main}
    \begin{algorithmic}
   \STATE {\bfseries Parameter:} The number of moments $G$.
   \STATE Obtain trained weights $\widehat{\bm{w}}^{\mathrm{MMSCM}}$ as \eqref{eq:gmm}. 
   \STATE Estimate the treatment effect as    \eqref{eq:att_est}. 
\end{algorithmic}
\end{algorithm}

\subsection{Convergence Analysis}
\label{sec:conv_analysis}
We now provide consistency results for our proposed SC weight estimator. The following theorem presents these results, including consistency for the treatment effect~$\tau_0$. 
\begin{assumption}
\label{asm:mixture2}
    The stochastic process $\{\varepsilon_{j, t}\}_{t=1}^{T_0}$ is stationary, strongly mixing, with the sum of mixing coefficients bounded by $M$.
\end{assumption}

\begin{theorem}
\label{thm:convergence}
Suppose that Assumptions~\ref{asm:bounded}--\ref{asm:core_asum} and \ref{asm:mixture2} hold. Let $v_\gamma \in (0, \infty)$ for $\gamma = 1, 2, \dots, G$ be a fixed weight sequence. Define
\begin{align*}
    \widetilde{\Phi}^\dagger \coloneqq \argmin_{\bm{w}\in\Delta^J}
    \sum_{\gamma\in\{1,2,\dots, G\}} 
    v_\gamma
    \left|\mathbb{E}_{0,t}\left[\left(Y^N_{0,t}\right)^\gamma \right] 
    - \sum_{j\in\mathcal{J}^U} w_j \mathbb{E}_{j,t}\left[\left(Y^N_{j,t}\right)^\gamma\right]\right|.
\end{align*}
Then, it holds that 
\[
\widehat{\bm{w}}^{\mathrm{MMSCM}}_{T_0, G} \xrightarrow{\mathrm{p}} \bm{w}^*_G \in \widetilde{\Phi}^\dagger
\quad \text{as } T_0\to \infty.
\]
Additionally, $\bm{w}^*_G \to \bm{w}^*\in \widetilde{\Phi}^*$ as $G\to \infty$, and 
\[
\widehat{\tau}^{\mathrm{MMSCM}}_{0, t} \xrightarrow{\mathrm{p}} \tau_{0,t} 
+ \left(\epsilon_{0, t} - \sum_{j\in\mathcal{J}^U}w^*_{j} \epsilon_{j, t}\right)
\quad \text{as } T_0 \to \infty \text{ and } G\to \infty.
\]
\end{theorem}
The proof is shown in Appendix~\ref{appdx:thm:convergence}. Note that this result also applies to $\widehat{\tau}^{\mathrm{MMSCM}}_{0, t}$ as a special case. 

Consequently, because $\mathbb{E}\left[\epsilon_{0, t} - \sum_{j\in\mathcal{J}^U}w^*_{j} \epsilon_{j, t}\right] = 0$, $\widehat{\tau}^{\mathrm{MMSCM}}_{0, t}$ is asymptotically unbiased.

\textbf{Relationship to the classical SCMs.} We next consider the relationship between our MMSCM and the classical SCMs proposed in \citet{AbadieGardeazabal2003}. \citet{AbadieGardeazabal2003} assumes that $Y^N_{0,t} = \sum_{j\in[J]} w^*_j Y^N_{j, t}$, which corresponds to a specific case of our mixture model assumption, $F_{Y^N_{0,t}}(y) = \sum_{j\in\mathcal{J}^U}w^*_j F_{Y^N_{j,t}}(y)\quad \forall y \in \mathbb{R}$. Therefore, under the assumptions in \citet{AbadieGardeazabal2003}, by using the MMSCMs, we can estimate the weights in an asymptotically unbiased way.

\begin{corollary}
    Suppose that Assumptions~\ref{asm:bounded} and \ref{asm:mixture2} hold. Assume that 
    \[Y^N_{0,t} = \sum_{j\in[J]} w^*_j Y^N_{j, t}\]
    holds. Let $v_\gamma \in (0, \infty)$ for $\gamma = 1, 2, \dots, G$ be a fixed weight sequence. Define
\begin{align*}
    \widetilde{\Phi}^\dagger \coloneqq \argmin_{\bm{w}\in\Delta^J}
    \sum_{\gamma\in\{1,2,\dots, G\}} 
    v_\gamma
    \left|\mathbb{E}_{0,t}\left[\left(Y^N_{0,t}\right)^\gamma \right] 
    - \sum_{j\in\mathcal{J}^U} w_j \mathbb{E}_{j,t}\left[\left(Y^N_{j,t}\right)^\gamma\right]\right|.
\end{align*}
Then, it holds that 
\[
\widehat{\bm{w}}^{\mathrm{MMSCM}}_{T_0, G} \xrightarrow{\mathrm{p}} \bm{w}^*_G \in \widetilde{\Phi}^\dagger
\quad \text{as } T_0\to \infty.
\]
Additionally, $\bm{w}^*_G \to \bm{w}^*\in \widetilde{\Phi}^*$ as $G\to \infty$, and 
\[
\widehat{\tau}^{\mathrm{MMSCM}}_{0, t} \xrightarrow{\mathrm{p}} \tau_{0,t} 
+ \left(\epsilon_{0, t} - \sum_{j\in\mathcal{J}^U}w^*_{j} \epsilon_{j, t}\right)
\quad \text{as } T_0 \to \infty \text{ and } G\to \infty.
\]
\end{corollary}

\begin{remark}[Section~4.1 in \citet{Gunsilius2020}]
    The proposed method is an extension of the classical method in a rigorous sense: if
we apply it to probability measures supported on one point, that is, Dirac measures of
the form $\delta_y(A)$, taking the value $1$ if $y \in A$ and $0$ otherwise, then we obtain the same
results as the classical method. This means the proposed method reduces to the classical estimator when we are only given aggregate values and not distributions.
\end{remark}
\color{black}

\subsection{Inference}
\label{sec:inference}
This section presents methods for statistical inference using our proposed MMSCM estimator. We consider statistical inference for the treatment effect using our proposed MMSCM and D2SCM, focusing on testing a sharp null hypothesis, $H_0 : \tau_{0, t} = \alpha_t$ for some $t \in \mathcal{T}_t$ and $\bm{\alpha} = (\alpha_{t})_{t\in\mathcal{T}_1}$. Specifically, we adopt the conformal inference approach of \citet{Chernozhukov2021}, which has been used by, for example, \citet{Ben2021} and \citet{Ferman2021}. \mkato{Note that under our linear factor model, $\tau_{0,t} = Y^N_{0,t} - Y^I_{0,t} = \alpha_t$ holds, that is, the ATE equals the individual treatment effect. Therefore, our null is sharp. This formulation is also adopted in other works, such as \citet{Chernozhukov2021}.}

Our goal is to construct a confidence interval whose $p$-value $\widehat{p}$ satisfies
\begin{align}
\label{eq:p_conv}
    \mathrm{Pr}\left(\widehat{p} \leq a\right) \to a \quad \text{for all } a \in(0, 1)
\quad \text{as } T_0 \to \infty \text{ and } T_1 \text{ is fixed}.
\end{align}
To achieve this, conformal inference proceeds as follows. First, for the sharp null hypothesis 
\[
H_0 : \tau_{0, t} = \alpha_t,
\]
we create adjusted post-treatment outcomes for the treated unit $Y^{N\sharp}_{0, t} = Y^I_{0, t} - \alpha_t$ and combine them with the original dataset as 
\[
\left(Y_{0, 1}, \dots, Y_{0, T_0}, Y^{N\sharp}_{0, (T_0 + 1)}, \dots,  Y^{N\sharp}_{0, T}\right).
\]

We then apply our proposed method to this augmented dataset, comprising 
$
\left(Y_{0, 1}, \dots, Y^{N\sharp}_{0, (T_0 + 1)}, \dots,  Y^{N\sharp}_{0, T}\right)
$
and $(Y_{j, 1})_{t\in\mathcal{T}, j\in\mathcal{J}^U}$, to obtain adjusted weights $\widehat{\bm{w}}(\alpha)$ and an adjusted predictor of the counterfactual outcome, denoted $\widehat{Y}^N_{0, t}(\bm{\alpha})$. 

Finally, we compute the $p$-value by assessing whether the adjusted residual ``conforms'' with the pre-treatment residuals. Define $u_t(\bm{\alpha}) = Y_{0, t} - \widehat{Y}^N_{0, t}(\bm{\alpha})$ for $t\in\mathcal{T}_0$, and $u_t(\bm{\alpha}) = Y^I_{0, t} - \alpha_t - \widehat{Y}^N_{0, t}(\bm{\alpha})$ for $t\in\mathcal{T}_1$. For $j\in \mathcal{T}$, let $\pi_j$ be a permutation of indices defined by $\pi_j(t) = t + j$ if $t + j \leq T$, and $\pi_j(t) = t + j - T$ otherwise. Let $\Pi = \{\pi_{1}, \dots, \pi_{T - 1}\}$. Then the $p$-value is defined as $\widehat{p}(\bm{\alpha}) = 1 - \widehat{F}(\bm{\alpha})$, where 
\[
\widehat{F}(\bm{\alpha}) = \frac{1}{T}\sum_{j\in\mathcal{T}} \mathbbm{1}\left[
\frac{1}{T}\sum_{t\in\mathcal{T}}\left| u_t(\bm{\alpha}) \right|
< \frac{1}{T}\sum_{\pi_{j}\in\mathcal{T}}\left| u_{\pi_{j}}(\bm{\alpha}) \right|\right].
\]
Because $Y^N_{0, t}$ is random, inverting this test yields a confidence interval for $\tau_{0, t}$, which is equivalent to constructing a conformal prediction set \citep{Vovk2005} for $Y^N_{0, T}$. The $(1-\xi)$ confidence interval is
\[
\widehat{C}^{\mathrm{conf}}_{1-\xi} = 
\left\{ \bm{\alpha}\in\mathcal{A}  \bigm| 
 \widehat{p}(\bm{\alpha}) > \xi\right\},
\]
where $\mathcal{A}$ is a set of candidate values. Theorem~1 of \citet{Chernozhukov2021} guarantees \eqref{eq:p_conv}. We summarize the procedure in Algorithm~\ref{alg:conformal}

\begin{remark}[Conformal inference and randomization inference]
    Conformal inference provides confidence intervals under a stationarity assumption and a prediction model for the counterfactual outcome. In contrast, randomization inference requires random assignment and prediction models for post treatment outcomes of all units. These differences are recognized advantages of conformal inference \citep{Chernozhukov2021}, and the method has recently been widely used in SCMs \citep{Ben2021}. Furthermore, conformal inference can itself be interpreted as a variant of randomization inference \citep{Ritzwoller2025randomizationinference}.
\end{remark}
\color{black}

\begin{algorithm}[t]
   \caption{Conformal prediction.}
   \label{alg:conformal}
    \begin{algorithmic}
   \STATE {\bfseries Sharp null hypothesis:} $H_0 : \tau_{0, t} = \alpha_t$ for some $t \in \mathcal{T}_t$ and $\bm{\alpha} = (\alpha_{t})_{t\in\mathcal{T}_1}$. 
   \STATE Make an augmented dataset $\left(Y_{0, 1}, \dots, Y_{0, T_0}, Y^{N\sharp}_{0, (T_0 + 1)}, \dots,  Y^{N\sharp}_{0, T}\right)$.
   \STATE  We apply the MMSCM by regarding the datasets
$
\left(Y_{0, 1}, \dots, Y^{N\sharp}_{0, (T_0 + 1)}, \dots,  Y^{N\sharp}_{0, T}\right)
$
and $(Y_{j, 1})_{t\in\mathcal{T}, j\in\mathcal{J}^U}$ as the pretreatment dataset. 
\STATE Using the estimated weight, we obtain estimate of the counterfactual outcome $\widehat{Y}^N_{0, t}(\bm{\alpha})$. 
\STATE Obtain $u_t(\bm{\alpha}) = Y_{0, t} - \widehat{Y}^N_{0, t}(\bm{\alpha})$ for $t\in\mathcal{T}_0$, and $u_t(\bm{\alpha}) = Y^I_{0, t} - \alpha_t - \widehat{Y}^N_{0, t}(\bm{\alpha})$ for $t\in\mathcal{T}_1$. \STATE Compute the $p$-value as $\widehat{p}(\bm{\alpha}) = 1 - \widehat{F}(\bm{\alpha})$ and confidence interval as $
\widehat{C}^{\mathrm{conf}}_{1-\xi} = 
\left\{ \bm{\alpha}\in\mathcal{A}  \bigm| 
 \widehat{p}(\bm{\alpha}) > \xi\right\}$. 
\end{algorithmic}
\end{algorithm} 

\subsection{Distributional Treatment Effect Estimation}
To estimate the distributional treatment effect, we approximate the counterfactual distribution $p_{0, t}(y)$ using a bootstrap-like simulation-based method. Specifically, we resample $L > 1$ observations $\{Y^{l, \dagger}_{0}\}_{l=1}^{L}$ from $\{Y_{0, t}, Y_{1, t}, \dots, Y_{J, t}\}_{t \in \mathcal{T}_1}$ as follows: for each $l \in \mathcal{L} \coloneqq \{1, 2, \dots, L\}$,
\begin{enumerate}
\item For each $j \in \mathcal{J}$, resample $Y^{l, \dagger}_{j}$ from $\{Y_{j, t}\}_{t \in \mathcal{T}_1}$.
\item Choose $Y^{l, \dagger}_{0}$ from $\left(Y^{l, \dagger}_{1}, \dots, Y^{l, \dagger}_{J}\right)$ with probability $\left(\widehat{w}^{\mathrm{MMSCM}}_{j, T_0}\right)_{j \in \mathcal{J}^U}$.
\end{enumerate}
The empirical distribution of $\{Y^{l, \dagger}_{0}\}_{l=1}^{L}$ serves as an estimator of the distribution of $\frac{1}{T_1} \sum_{t \in \mathcal{T}_1} Y_{0, t}$, as shown in the following theorem. The proof is provided in Appendix~\ref{appdx:distconv}. Thus, when our primary focus is on $\frac{1}{T_1} \sum_{t \in \mathcal{T}_1} p_{0, t}(y)$, the above procedure enables us to achieve this goal.

\begin{theorem}
\label{thm:distconv}
    Suppose that the assumptions in Theorem~\ref{thm:convergence} hold. Define 
    \begin{itemize}
        \item $G_{T_0, L} \coloneqq \frac{1}{L} \sum^L_{l=1} \mathbbm{1}[Y^{l, \dagger}_{0} \leq y]$.
        \item $F_{Y^N_{j,t}, T_0}(y) \coloneqq \frac{1}{T_0} \sum_{t \in \mathcal{T}_0} \mathbbm{1}[Y^N_{j, t} \leq y]$.
        \item $\widehat{F}_{Y^N_{0,t}, T_0}(y) \coloneqq \sum_{j \in \mathcal{J}^U} \widehat{w}^{\mathrm{MMSCM}}_{j, T_0, G} F_{Y^N_{j,t}, T_0}(y)$.
    \end{itemize}
    Then, the following holds:
    \begin{align*}
        \sup_{y} \big|G_{T_0, L}(y) - F_{Y^N_{0,t}}(y)\big| \xrightarrow{\mathrm{p}} 0 \quad (L \to \infty, T_0 \to \infty),
    \end{align*}
    where we recall that $F_{Y^N_{0,t}}(y) = \sum_{j \in \mathcal{J}^U} w^*_j F_{Y^N_{j,t}}(y) \quad \forall y \in \mathbb{R}$.
\end{theorem}

\textbf{Two-sample homogeneity test.}
To test distributional treatment effects, consider the null and alternative hypotheses $H_0: p^I_t = p^N_t$ versus $H_1: p^I_t \neq p^N_t$. One can perform a two-sample homogeneity test in combination with our distributional treatment effect estimator. For instance, one may employ maximum mean discrepancy (MMD) \citep{Gretton2012} to implement this hypothesis test.

\color{black}

\section{Discussion}
\label{sec:discuss}
We conclude by discussing some remaining issues.

\subsection{Justification of Mixture Models}
\label{sec:just_mm}
To clarify assumptions in SCMs, \citet{Shi2022} proposes fine-grained models wherein each unit is composed of multiple components, and the outcomes of treated and untreated units are averages of the outcomes across these components. For example, in the empirical study of California’s tobacco control program, the outcome of interest (annual per capita cigarette consumption) at the state level is an average of individuals’ consumption in the state. Building on this framework, one can justify mixture model assumptions by relating them to linear factor models.

Assume there are $U_j$ components in unit $j \in \mathcal{J}$. For each $u\in \mathcal{U}_j \coloneqq \{1,2,\dots,U_j\}$, let $(Y^I_{j, t, u}, Y^N_{j, t, u})$ be the potential outcomes of component $u$ in unit $j$ at time $t$. Suppose there is a $d_W$-dimensional unobserved random variable $W_{j, t, u} \in \mathbb{Z}^{d_W}$ capturing individual characteristics (e.g., age or income). We assume $(Y^I_{j, t, u}, Y^N_{j, t, u}, W_{j,t,u})$ is i.i.d.\ across $u \in \mathcal{U}_j$. 

The relationship between $W_{j, t, u}$ and potential outcomes $(Y^I_{j, t, u}, Y^N_{j, t, u})$ can be linear or nonlinear, and possibly time-varying. Let $p_{j,t}(w, y)$ be the PDF of $(W_{j,t}, Y^N_{j,t})$, and let $p_{j,t}(y\mid w)$ be the conditional PDF of $Y^N_{j,t}$ given $W_{j,t}$, while $p_{j, t}$ is the marginal PDF of $W_{j, t}$. We assume that the intervention is administered at a group level and that individuals within each group comply with their group-level treatment.

Under these conditions, \citet{Shi2022} shows that the following assumptions are required for the linearity of SCMs in \eqref{eq:linear_exp}:

\begin{assumption}[Independent Causal Mechanism; ICM]
\label{asm:icm}
Conditional on the causes $W_{j, t, u}$, the potential outcome $Y^N_{j, t, u}$ is independent of the unit $j \in \mathcal{J}$. For the population distribution $j \in \mathcal{J}$ at time $t \leq \mathcal{T}$, the joint distribution of $(Y^N_{j, t, u}, W_{j, t, u})$ is given by $p_{j, t}(w, y) = r_{j,t}(w) p_t(y\mid w)$.
\end{assumption}

\begin{assumption}[Stable distributions]
\label{asm:stable}
Decompose the causes as $W_{j, t, u} = \{U_{j, t, u}, S_{j, t, u}\}$. Let $S$ denote the subset of causes that differentiates the treated unit from the selected untreated units (the minimal invariant set). We assume that, for all groups, the distribution of $S$ does not change for all time periods $t \leq T$. Hence, $r_{j, t}(w) = \zeta_j(s) \xi_t(u\mid s)$, where $\zeta_j(s)$ is the PDF of $S_{j, t, u}$, and $\xi_t(u\mid s)$ is the PDF of $U_{j, t, u}$ given $S_{j, t, u}$.
\end{assumption}

\begin{assumption}[Sufficiently similar untreated units]
\label{asm:sim}
Let $S$ be the minimal invariant set for the target and untreated units. The untreated units are sufficiently similar if the cardinality of this minimal invariant set satisfies $\left|\mathcal{J}^U\right| \geq \lvert S\rvert$.
\end{assumption}

\begin{assumption}[Common support]
\label{asm:common}
Let $(s_1, \dots, s_R)$ be the support of $S$. For each $s$, there is at least one untreated unit $j$ with $\zeta_j(S=s) > 0$.
\end{assumption}

\begin{proposition}[Causal identifiability. From Theorem~1 of \citet{Shi2022}]
\label{prp:shi}
Suppose Assumptions~\ref{asm:icm}--\ref{asm:common} hold. Then there exist weights $\{w^*_j\}_{j\in[J]}$ such that $p^N_{0, t}(y) = \sum_{j\in\mathcal{J}^U} w^*_j  p^N_{j, t}(y)$ holds.
\end{proposition}

This result also implies that the mixture model \eqref{eq:mixture_model} holds \citep{nazaret2023misspecification}, aligning it with linear factor models. Hence, we obtain the following corollary for linear factor models similar to \citet{Ferman2021}:

\begin{corollary}
\label{cor:conv}
Suppose that Assumptions~\ref{asm:bounded}, \ref{asm:mixture2}, and \ref{asm:icm}--\ref{asm:common} hold. Then 
\[
\bm{w}^{\mathrm{MMSCM}}_{T_0, G} \xrightarrow{\mathrm{p}} \bm{w}^*_G
\quad \text{as } T_0 \to \infty,
\]
and $\widehat{\tau}^{\mathrm{MMSCM}}_{0, t}$ converges to 
\[
\tau_{0,t} + \left(\epsilon_{0, t} 
  - \sum_{j\in\mathcal{J}^U}w^*_{j} \epsilon_{j, t}\right)
\]
in probability as $T_0 \to \infty$ and $G\to\infty$.
\end{corollary}

\noindent
\emph{Proof Sketch.} From Proposition~\ref{prp:shi}, under Assumptions~\ref{asm:icm}--\ref{asm:common}, \eqref{eq:mixture_model} holds. Therefore, applying Theorem~\ref{thm:convergence} yields the desired result under \eqref{eq:latent} with $c_j = 0$ for each $j\in\mathcal{J}$.

\subsection{Auxiliary Covariates}
We can also extend our method to incorporate auxiliary covariates if they are observable. For each unit $j\in \mathcal{J}$, let
\[
X_{j,t}\coloneqq (X_{j, t, 1}, X_{j, t, 2}, \dots, X_{j, t, K})\in\mathbb{R}^{K}
\]
be $K$-dimensional covariates that are unaffected by treatment. Define
\[
\widehat{\ell}_{k}(\bm{w}) 
= \frac{1}{T_0}\sum_{t\in\mathcal{T}_0}\left(X_{0, t, k} 
- \sum_{j\in\mathcal{J}^U} w_j  X_{j, t, k}\right).
\]
When these covariates are available, we can extend the MMSCM estimator as
\mkato{
\[
\widehat{\bm{w}}^{\mathrm{MMSCM}}_{T_0}
\coloneqq \argmin_{\bm{w}\in\Delta^J}
b_0\sum_{\gamma=1}^G 
v_\gamma
\left|\frac{1}{T_0}\sum_{t\in\mathcal{T}_0}
\left(\left(Y^N_{0,t}\right)^\gamma  - \sum_{j\in\mathcal{J}^U} w_j \left(Y^N_{j,t}\right)^\gamma\right)\right|
+\sum_{k=1}^K b_k  \widehat{\ell}^2_{k}(\bm{w}),
\]
}
where $(b_k)_{k=0}^K$ are weights such that $b_k \geq 0$ and $\sum_{k=0}^K b_k = 1$.

\subsection{Distributional SCMs and Optimal Transport}
\citet{Gunsilius2020} propose \emph{distributional SCMs}, which assume
\[
F^{-1}_{Y_{0, t}, N}(q) 
= \sum_{j\in\mathcal{J}^U}w^*_j F^{-1}_{Y_{j, t}, N}(q)
\quad \forall q \in (0, 1),
\]
where $F^{-1}_{Y_{j, t}, N}$ is the quantile function defined by
\[
F^{-1}_{Y_{j, t}, N}(q) 
\coloneqq \inf\left\{ y\in\mathbb{R}\colon F_{Y_{j, t}, N}(y) \geq q\right\}.
\]
To estimate $\bm{w}^*$, \citet{Gunsilius2020} minimize the Wasserstein distance, referring to their method as DiSCo. Concretely, the DiSCo estimator is
\begin{align*}
    \widehat{\bm{w}}^{\mathrm{DiSCo}} 
    \coloneqq 
    \argmin_{\bm{w}\in \Delta^J}
    \frac{1}{M}\sum_{m=1}^M 
    \left|
    \widehat{F}^{-1}_{Y^N_{0, t}}(V_m) 
    - 
    \sum_{j\in\mathcal{J}^U}w_j 
    \widehat{F}^{-1}_{Y^N_{j, t}}(V_m)
    \right|,
\end{align*}
where $\widehat{F}^{-1}_{Y^N_{j, t}}(V_m)$ is the empirical quantile function of $Y^N_{j, t}$, and $\{V_m\}_{m=1}^M$ are i.i.d.\ draws from the uniform distribution on $[0, 1]$.

\begin{figure}[t]
    \centering
    \includegraphics[width=0.9\linewidth]{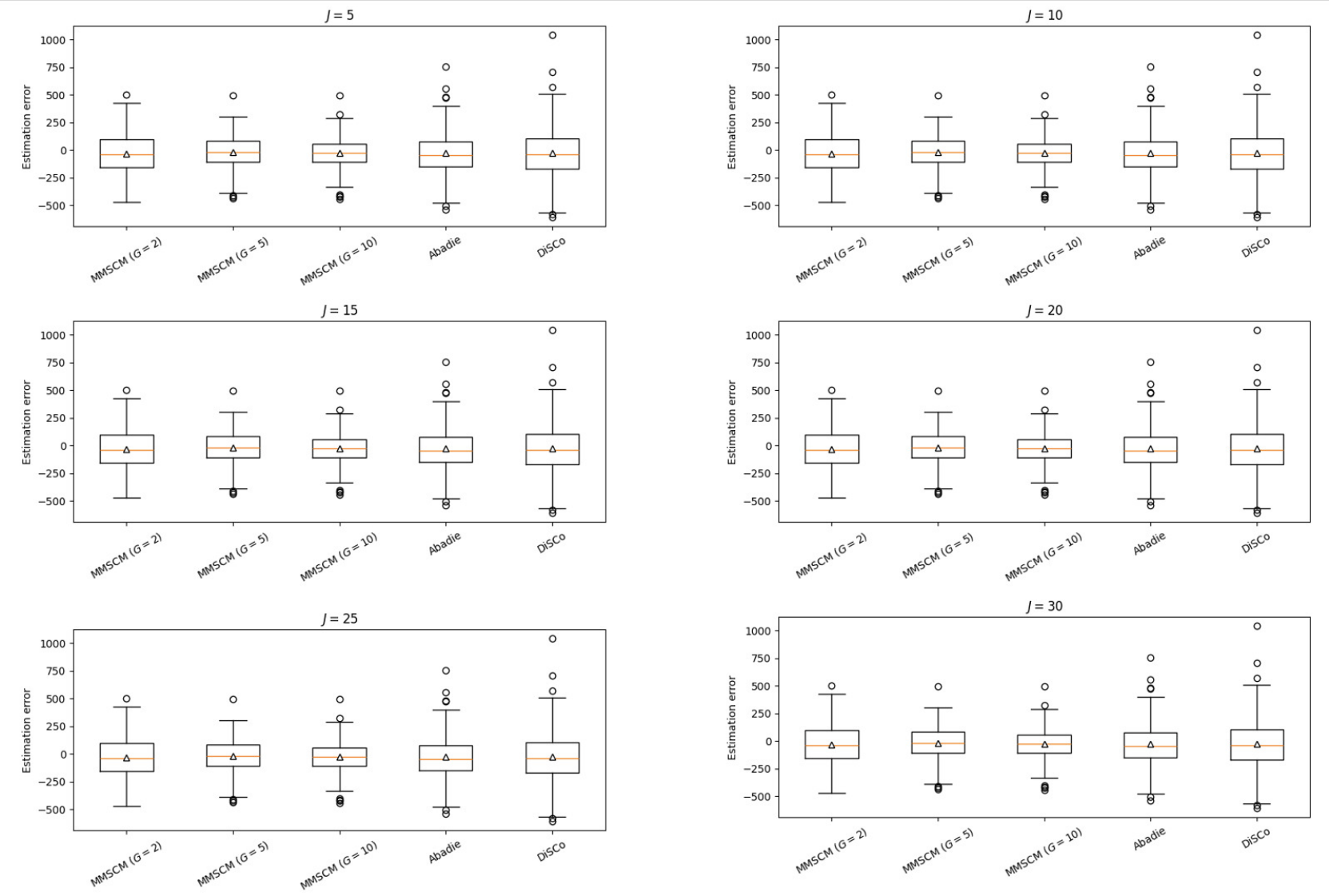}
    \caption{Results of simulation studies. Each result represents estimation error ($\tau_{0, t} - \widehat{\tau}$ for an estimator $\widehat{\tau}$). The orange horizontal line denotes the median, and the triangle mark denotes the mean.}
    \label{fig:num_exp}
\end{figure}

\begin{figure}[t]
    \includegraphics[width=0.9\linewidth]{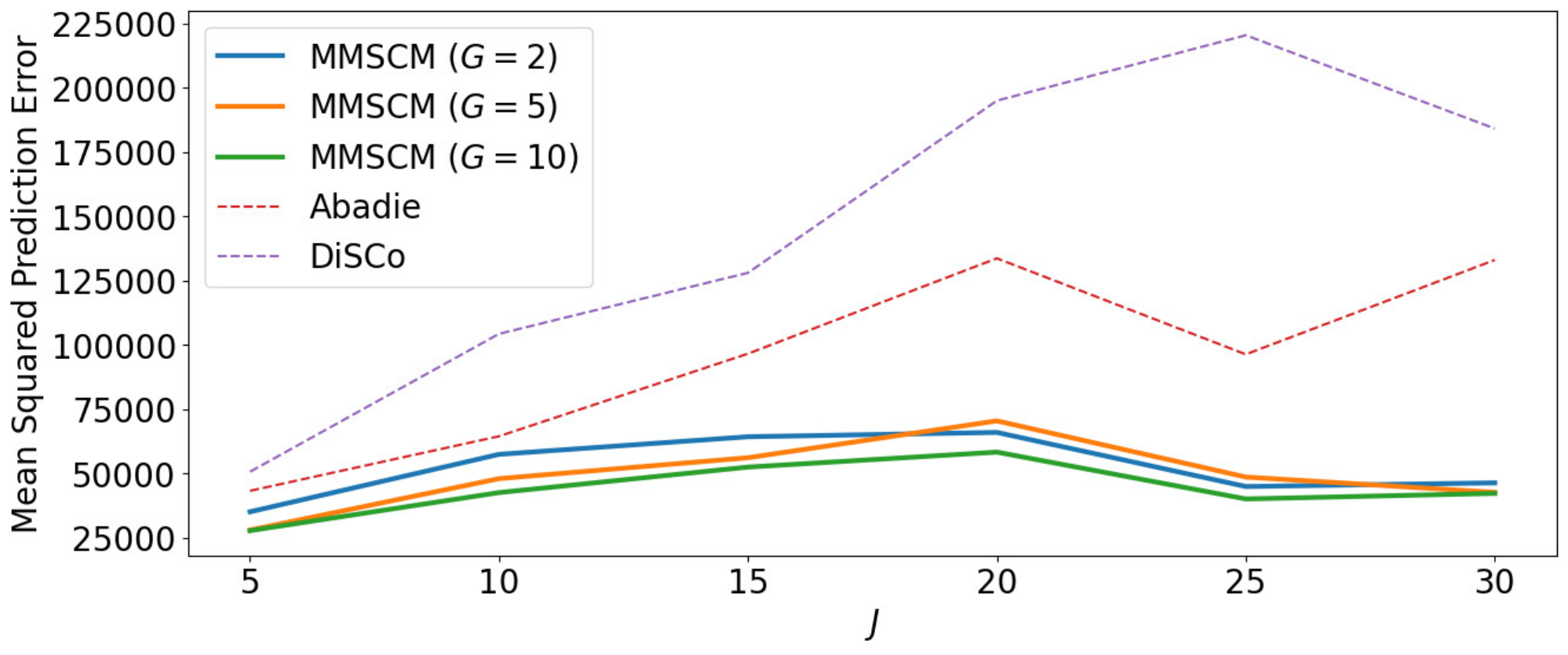}
    \caption{Results of simulation studies on ATE prediction. The $x$-axis is the number of untreated units $J$, and the $y$-axis is the mean squared prediction error.}
    \label{fig:doubledescent}
\end{figure}
\begin{figure}[t]
    \includegraphics[width=0.9\linewidth]{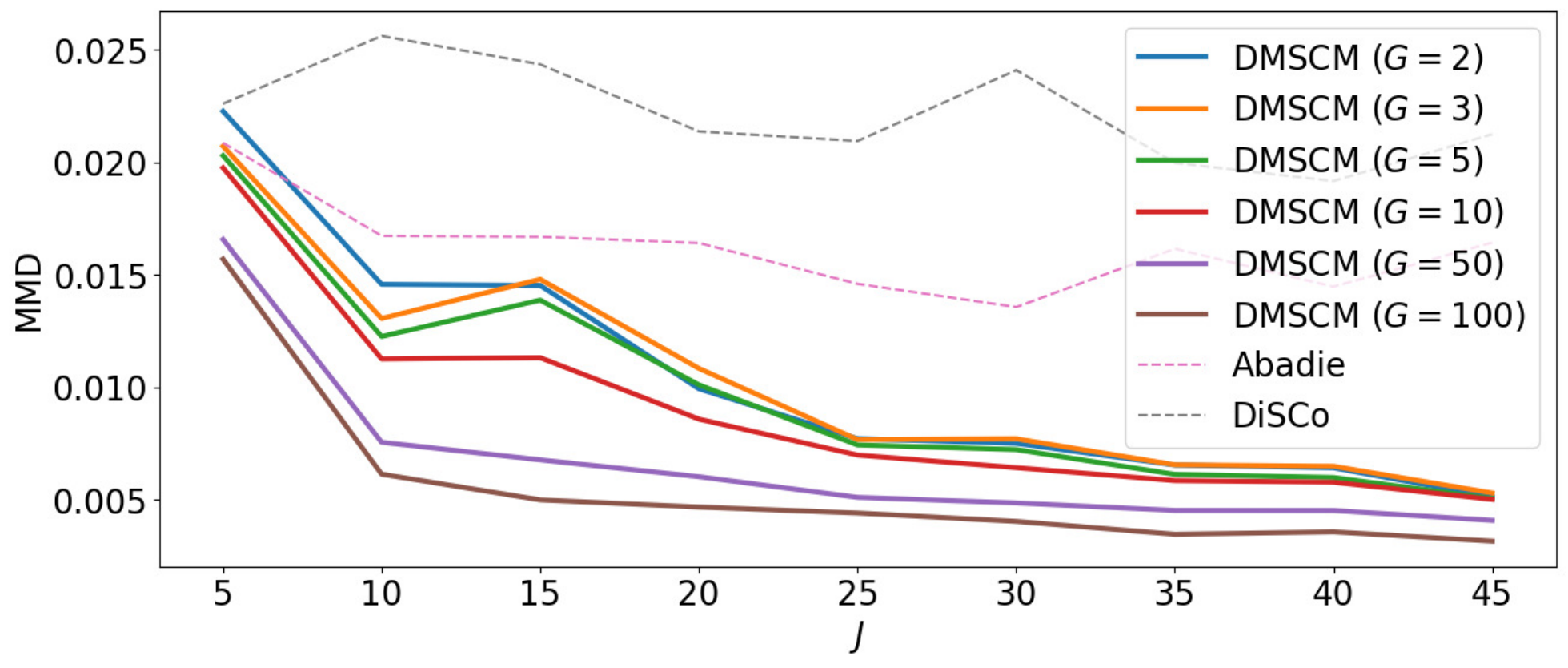}
    \caption{Results of simulation studies on distributional treatment effect prediction. The $x$-axis is the number of untreated units $J$, and the $y$-axis is the MMD.}
    \label{fig:doubledescent2}
\end{figure}

\section{Experiments}
\label{sec:exp}
We conduct experiments using both simulated and real-world data. In all experiments, we set $v_\gamma = 1$ for the MMSCM. 

\subsection{Simulation Studies}
\label{sec:sim_exp}
We first perform simulation studies to compare our proposed SCMs with traditional SCMs. Set $J\in\{10, 30, 60\}$ and $G \in \{2, 3, 5, 10, 20, 30, 40, 50, 60, 70, 80, 90, 100\}$, $T_0 = 30$, $T_1 = 100$, and $K=5$. 

We consider SCMs for 
\[
p_{0, t}(\bm{z}) = \sum_{j\in\mathcal{J}^U}w^*_j   p_{j, t}(\bm{z}),
\]
where $\bm{z}$ is a $(K+1)$-dimensional (i.e., $6$-dimensional) vector: the first element corresponds to $Y_{j, t}$ and the remaining elements correspond to $X_{j, t}$. For each $j\in \mathcal{J}$, let $p_{j, t}$ be the density of a multivariate normal distribution 
\[
\mathcal{N}\left(\bm{\mu}_{j, t}, \bm{\sigma}^2_{j, t}\right),
\]
where $\bm{\mu}_{j, t} = (\mu_{j, t, 1},\dots,\mu_{j, t, 6})$ and $\bm{\sigma}^2_{j, t}$ is a $(6\times 6)$ diagonal matrix whose $(k,k)$-element is $\sigma^2_{j, t, k}$. Each $\mu_{j, t, k}$ is drawn from a standard normal distribution, and each $\sigma^2_{j, t, k}$ is drawn from a uniform distribution on $[1, 20]$. For each $j\in\mathcal{J}$, we modify $\mu_{j, t, 1}$ and $\sigma^2_{j, t, 1}$ by adding noise: 
\[
\mu_{j, t, 1} \leftarrow \mu_{j, t, 1} + \epsilon_{j, t}, 
\quad
\sigma^2_{j, t, 1} \leftarrow \sigma^2_{j, t, 1} + \widetilde{\epsilon}_{j, t},
\]
where $\epsilon_{j, t}$ and $\widetilde{\epsilon}_{j, t}$ follow $\mathcal{N}(0, 10)$. If $\widetilde{\epsilon}_{j, t}$ is less than or equal to $0.1$, we set $\widetilde{\epsilon}_{j, t} = 0.1$ to avoid near-zero variance.

For the treated unit, we let $Y^N_{0, t} \sim p_{0, t}(\bm{z})$ for $t\in[t_0, T_1]$ and set $Y^I_{0, t} = 20 + Y^N_{0, t}$ for $t\in\mathcal{T}_1$, implying $\tau_{0, t} = 20$. We draw $\bm{w}^*$ from a uniform distribution over $[0,1]^J$ and then normalize it so that its entries sum to $1$.

We apply MMSCM, the method proposed in \citet{AbadieDiamondHainmueller2010} (referred to here as Abadie), and DiSCo in \citet{Gunsilius2020}. We iterate trials $100$ times and evaluate the estimation error of the treatment effect. We plot the estimation error box plots in Figure~\ref{fig:num_exp} \mkato{with medians and means}, indicate that our proposed method achieves smaller estimation errors. Moreover, the estimation error declines as $G$ increases.

Next, we consider another set of simulation studies under conditions similar to those above, but with the following changes: $J\in\{1, 5, 10, 15, 20, 25, 30\}$, $G \in \{2, 5, 10\}$, $T_0 = 30$, $T_1 = 1000$, and $K=5$. All other settings remain as in Section~\ref{sec:sim_exp}.

We again apply MMSCM and Abadie $100$ times and compute the error of the treatment effect estimation (Figure~\ref{fig:doubledescent}). We focus on how the mean squared estimation error changes as $J$ grows. While the estimation error under the method of \citet{Abadie2002} grows with $J$, our proposed method does not exhibit such growth. Additionally, our method maintains lower estimation errors than the method of \citet{Abadie2002}.

A similar pattern appears in distributional treatment effect estimation. We measure the discrepancy between the true and estimated distributions using the Maximum Mean Discrepancy \citep[MMD,][]{Gretton2012}, a probability metric based on kernel embeddings. In the top panel of Figure~\ref{fig:doubledescent2}, we show how MMD varies with $J\in\{1, 5, 10, 15, 20,\dots,35, 40, 45\}$. As before, our proposed methods outperform the method of \citet{Abadie2002} in this setting. In the bottom panel of Figure~\ref{fig:doubledescent2}, we plot MMD for $G \in \{2, 10, 50, 100\}$. As $G$ grows (i.e., as more moment conditions are imposed), the predictive performance improves. \mkato{Note that in this experiment, we included the results of \citet{AbadieDiamondHainmueller2010} for reference only, since it is not a method for distributional SCM. Our focus is on the gains induced by choosing a larger $G$.}

Our observations may connect with findings by \citet{spiess2023double}, who examine high-dimensional SCMs. Exploring these connections will be part of our future research.

\begin{figure}[h]
    \centering
            \includegraphics[width=0.8\linewidth]{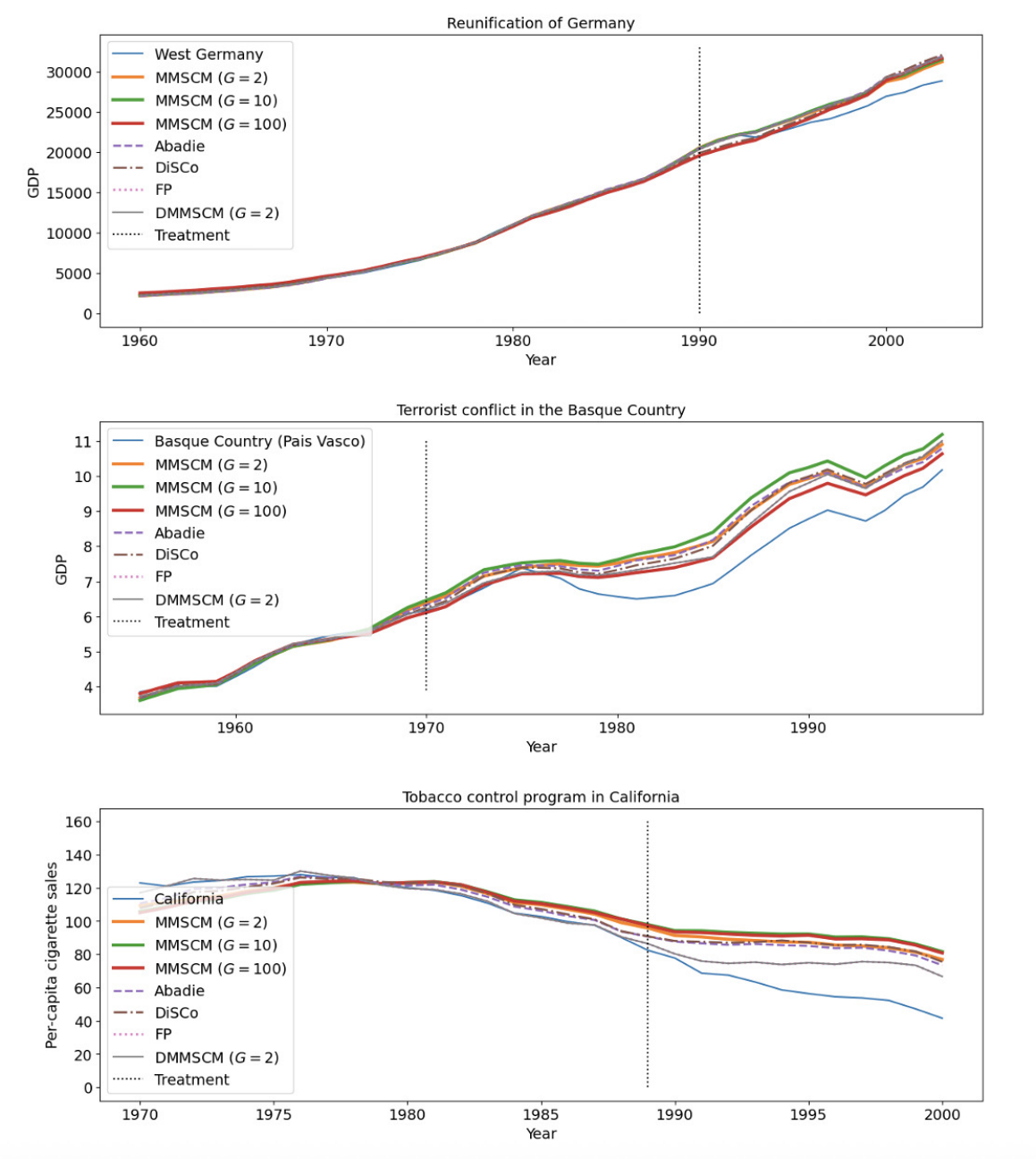}
        \caption{Counterfactual outcomes in empirical analysis.}
\label{fig:synthetic_results}
\end{figure}
\begin{figure}[h]
    \centering
        \includegraphics[width=0.8\linewidth]{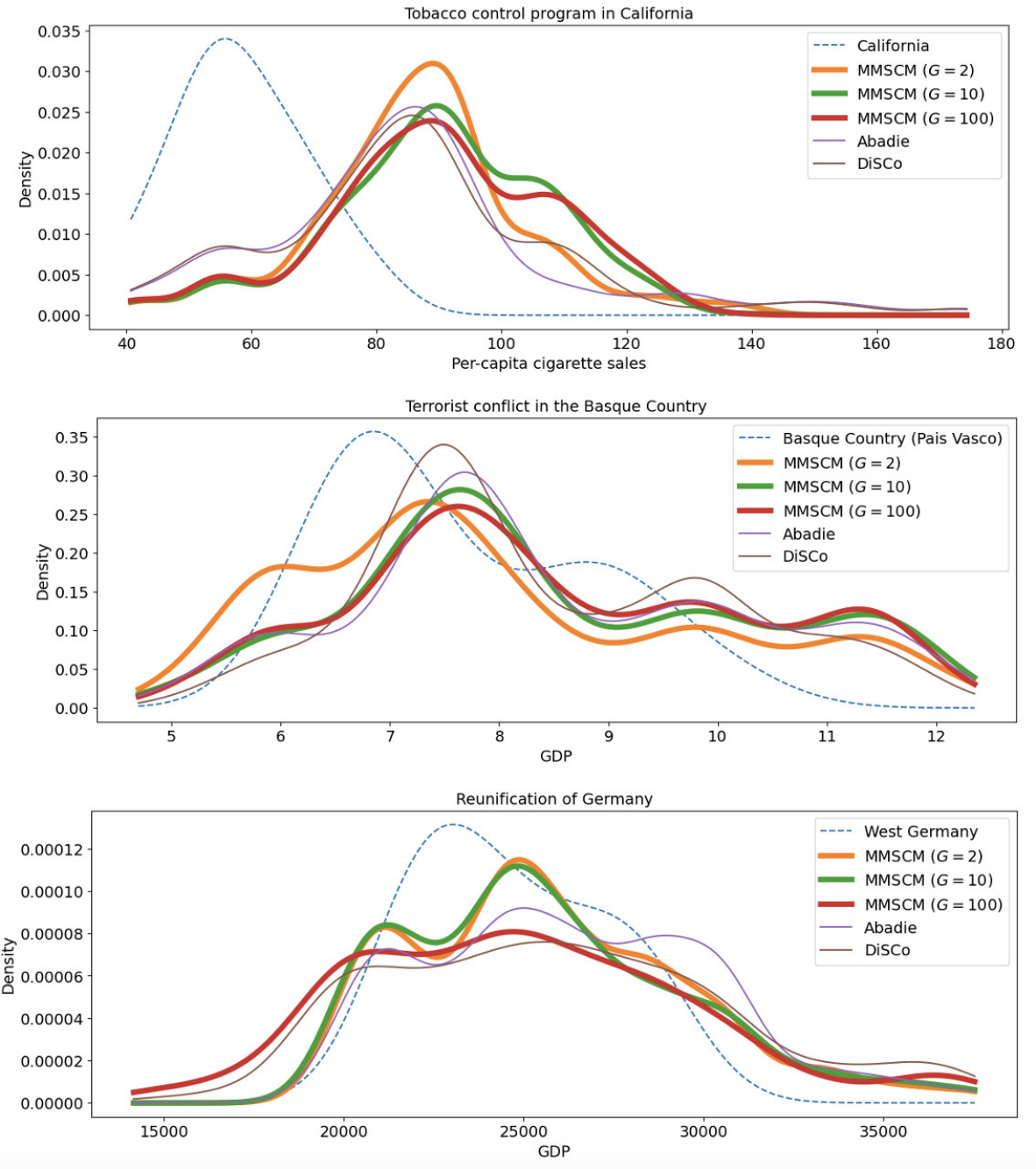}
        \caption{Counterfactual distributions in empirical analysis.}
\label{fig:synthetic_results_dist}
\end{figure}

\subsection{Empirical Analysis}
\label{sec:empanalysis}
We now present empirical studies based on the datasets and settings in \citet{AbadieGardeazabal2003}, \citet{AbadieDiamondHainmueller2010}, and \citet{AbadieDiamondHainmueller2015}. In this analysis, we also compare our method with the prediction intervals for SCMs proposed by \citet{CattaneoFeng2021}, which provide a robust uncertainty evaluation for SCMs. Although this prediction interval cannot be applied to our method due to differences in the underlying models, it serves as a promising alternative to conformal prediction. Therefore, we include its results for reference. Note that while we used our own implementation for our method, Abadie’s method, and DiSCo, we used the \texttt{pcsi} library for the prediction intervals, as published by \citet{Cattaneo2022scpiuncertainty}.

\textbf{Terrorist conflict in the Basque Country.}
\citet{AbadieGardeazabal2003} investigate the effect of terrorist conflict in the Basque Country on gross domestic product (GDP). In this example, $J = 16$, $K = 14$, and $(t_0, T_0, T_1) = (1955, 1970, 1997)$. Basque Euskadi Ta Askatasuna (Basque Homeland and Liberty; ETA) was founded in 1959 as a Basque separatist organization in Spain, seeking an independent Basque state primarily through terrorist activities in the Basque Country region. \citet{AbadieGardeazabal2003} use annual Spanish regional-level panel data from 1955 to 1997, with a particular focus on the 1970s when ETA’s terrorist activity was significant. The dataset includes fifteen years of pre-intervention observations. The set of untreated units consists of $17$ other Spanish regions. Each region’s record contains the outcome of interest (GDP per capita) and additional covariates (investment rate, population density, five sectoral output measures, and four measures of human capital), all averaged from 1955 to 1968.

\textbf{Tobacco control program in California.}
\citet{AbadieDiamondHainmueller2010} also investigate the causal effect of Proposition~99, a comprehensive tobacco control program adopted by California. They use annual state-level panel data from 1970 to 2000. The program was implemented in January 1989, providing eighteen years of pre-intervention observations. Four states that adopted large-scale tobacco control programs (Massachusetts, Arizona, Oregon, and Florida) and states that raised cigarette taxes by at least \$0.50 between 1989 and 2000 (Alaska, Hawaii, Maryland, Michigan, New Jersey, New York, and Washington) as well as the District of Columbia are excluded from the set of 38 control states. The outcome of interest is annual per capita cigarette sales in packs; covariates include the average retail price of cigarettes, log of state income per capita, the percentage of the population aged 15--24, and per capita beer consumption, all averaged between 1980 and 1988. Here, $J = 20$, $K = 5$, and $(t_0, T_0, T_1) = (1970, 1989, 2000)$. 

\textbf{Reunification of Germany.}
\citet{AbadieDiamondHainmueller2015} examine the impact of Germany’s reunification in 1990. Because no single country provided a suitable comparison, SCMs were used to create a composite unit. In this example, $J = 8$, $K = 31$, and $(t_0, T_0, T_1) = (1960, 1990, 2003)$. 

\textbf{Results.}
For each dataset, Figures~\ref{fig:synthetic_results} and \ref{fig:synthetic_results_dist} show the estimated treatment effects and the distributional treatment effects, respectively. In Appendix~\ref{appdx:confband},  Tables~\ref{tab:conformal_tabacco}--\ref{tab:conformal_reunification} report the estimated treatment effects and 90\% confidence bands obtained via conformal inference. We also plot the confidence bands in Figures~\ref{fig:confinterval_tabacco}--\ref{fig:confinterval_germany}.

Our proposed method fits the observed outcomes well in the pre-treatment period ($t \in \mathcal{T}_0$) and, accordingly, produces plausible counterfactual outcomes and distributions.

In the conformal inference analysis, when we set the sharp null hypothesis as $H_0: \tau_{0, (T_0 + 1)} = \cdots = \tau_{0, T_1} = 0$ for the treatment effect, none of the methods reject the null hypothesis, except for the ``California tobacco control'' dataset. Prediction intervals proposed in \citet{CattaneoFeng2021} shows tight intervals for every settings. Note that we cannot directly compare the performance of their method and ours due to differences in the underlying models.

\section{Conclusion}
This study investigated the asymptotically unbiased estimation of the treatment effect in the setting of SCMs. We first pointed out the problem of implicit endogeneity in the linear regression models commonly employed by existing SCMs, which can be interpreted as a form of measurement error bias. To address this issue, we focused on distributional SCMs, in which we assume mixture models for the outcome distributions and developed a moment-matching SCM. By matching the moments between the outcome of the treated unit and the weighted sum of the outcomes of the untreated units, our method consistently estimates SC weights and reinforces the theoretical foundation for using SCMs in comparative studies. We demonstrated the effectiveness of this approach through both simulation experiments and empirical applications.

\appendix

\section{Proof of Theorem~\ref{thm:convergence}}
\label{appdx:thm:convergence}
This section provides the proof for Theorem~\ref{thm:convergence}. Before presenting the proof, we introduce some existing results related to the moment problem and additional notations.

\textbf{Moment problem.}
The question of whether a sequence of moments uniquely determines a distribution is known as the moment problem \citep{Stchmudgen2020lecturesmomentproblem}. In particular, when the random variable is bounded, the problem is referred to as the Hausdorff moment problem. For this case, \citet{Hausdorff1921summationsmethodenund} provides the following result: if a random variable has support on $[0, 1]$, then the sequence of its moments is uniquely determined by its distribution.

\begin{proposition}
[\citet{Hausdorff1921summationsmethodenund}, Satz~III]
\label{prp:char}
Let $\{\mu_n\}_{n\ge0}$ be a real sequence.  
Define the forward differences recursively by
\[
  \Delta\mu_n \coloneqq \mu_{n+1}-\mu_n,\qquad
  \Delta^{k}\mu_n \coloneqq \Delta(\Delta^{k-1}\mu_n)\quad(k\ge1).
\]

The following are equivalent:
\begin{itemize}
  \item (\emph{Moment representation})  
        There exists a finite positive Borel measure $\nu$ supported on $[0,1]$ such that
        \[
          \mu_n = \int^1_0 u^{n}d\nu(u)\qquad(n=0,1,2,\dots).
        \]
  \item (\emph{Complete monotonicity})  
        The sequence is completely monotone, that is
        \[
          (-1)^{k}\Delta^{k}\mu_n \ge 0
          \qquad\text{for all }k,n\in\mathbb{N}.
        \]
\end{itemize}
Moreover, when these conditions hold, the representing measure $\nu$ is unique.
\end{proposition}
\color{black}

\textbf{Additional notations.}
Let us define
\begin{align*}
    &\widehat{Q}_G(\bm{w}) \coloneqq \sum_{\gamma\in\{1,2,\dots, G\}} v_\gamma\left|\frac{1}{T_0}\sum_{t\in\mathcal{T}_0}\left(\Big(Y^N_{j,t} \Big)^\gamma  - \sum_{j\in\mathcal{J}^U} w_j\Big(Y^N_{j,t} \Big)^\gamma\right)\right|,\\
    &Q_G(\bm{w}) \coloneqq \sum_{\gamma\in\{1,2,\dots, G\}} v_\gamma\left|\mathbb{E}_{0,t}\left[\Big(Y^N_{j,t} \Big)^\gamma \right] - \sum_{j\in\mathcal{J}^U} w_j\mathbb{E}_{j,t}\left[\Big(Y^N_{j,t} \Big)^\gamma\right]\right|.
\end{align*}

\textbf{Proof.}
Then, we prove Theorem~\ref{thm:convergence} as follows:
\begin{proof}
First, from the definition of $\Delta^J$, $\Delta^J$ is a compact parameter space. Next, $\widehat{Q}(\bm{w})$ converges uniformly in probability to $Q(\bm{w})$ as $T_0 \to \infty$ \mkato{under Assumptions~\ref{asm:bounded} and \ref{asm:mixture2} \citep{Potscher1989auniform}}. Furthermore, $Q(\bm{w})$ is continuous with respect to $\bm{w}$. 

Under these conditions, for any $\varepsilon > 0$, we have with probability approaching one, we have
\begin{enumerate}
    \item $\widehat{Q}_G\left(\widehat{\bm{w}}^{\mathrm{MMSCM}}\right) < \widehat{Q}(\bm{w}_0) + \varepsilon / 3$;
    \item \mkato{$Q_G(\widehat{\bm{w}}^{\mathrm{MMSCM}}) < \widehat{Q}_G\left(\widehat{\bm{w}}^{\mathrm{MMSCM}}\right) + \varepsilon / 3$};
    \item $\widehat{Q}_G\left(\bm{w}_0\right) < Q_G(\bm{w}_0) + \varepsilon / 3$.
\end{enumerate}
Therefore, with probability approaching one,
\begin{align*}
    Q_G\left(\widehat{\bm{w}}^{\mathrm{MMSCM}}\right) < \widehat{Q}_G\left(\widehat{\bm{w}}^{\mathrm{MMSCM}}\right) + \varepsilon / 3 < \widehat{Q}_G\left(\bm{w}_0\right) + 2\varepsilon / 3 < Q_G\left(\bm{w}_0\right) + \varepsilon.
\end{align*}
Thus, for any $\varepsilon > 0$, $Q_G\left(\widehat{\bm{w}}^{\mathrm{MMSCM}}\right) < Q_G\left(\bm{w}_0\right) + \varepsilon$ holds with probability approaching one. Let $\mathcal{N}$ be any open subset of $\Delta^J$ containing $\widetilde{\Phi}^*$. Since $\Delta^J \cap \mathcal{N}^c$ is compact, and $Q_G(\bm{w})$ is continuous with respect to $\bm{w}$, we have
\[\inf_{\bm{w} \in \Delta^J \cap \mathcal{N}^c}Q_G(\bm{w}) = Q_G(\bm{w}^\dagger) > Q_G(\bm{w}_0)\]
for some $\bm{w}^\dagger \in \Delta^J \cap \mathcal{N}^c$. Thus, choosing $\varepsilon = Q_G(\bm{w}_0) - \inf_{\bm{w} \in \Delta^J \cap \mathcal{N}^c}Q_G(\bm{w})$, it follows that
\begin{align*}
     Q_G\left(\widehat{\bm{w}}^{\mathrm{MMSCM}}\right) < \inf_{\bm{w} \in \Delta^J \cap \mathcal{N}^c}Q_G(\bm{w}),
\end{align*}
with probability approaching one. This implies that $\widehat{\bm{w}}^{\mathrm{MMSCM}} \notin \Delta^J \cap \mathcal{N}^c$, and hence $\widehat{\bm{w}}^{\mathrm{MMSCM}} \in \mathcal{N}$, where recall that  $\mathcal{N}$ is any open subset of $\Delta^J$ containing $\widetilde{\Phi}^*$. Therefore, $\widehat{\bm{w}}^{\mathrm{MMSCM}}_{T_0, G} \xrightarrow{\mathrm{p}} \bm{w}^*_G \in \widetilde{\Phi}^\dagger$ holds as $T_0\to \infty$. Additionally, from Proposition~\ref{prp:char}, we have $\bm{w}^*_G \to \bm{w}^*\in \widetilde{\Phi}^*$ as $G\to \infty$.
\end{proof}

\section{Proof of Theorem~\ref{thm:distconv}}
\label{appdx:distconv}
\begin{proof}
    Each draw $Y^{l, \dagger}_{0}$ is independent and distributed according to $\widehat{F}_{Y^N_{0,t}, T_0}(y)$. Therefore, for any fixed $T_0$, we have
    \[G_{T_0, L} \xrightarrow{\mathrm{a.s.}}\widehat{F}_{Y^N_{0,t}, T_0}(y)\]
    as $B \to \infty$, from the Glivenko–Cantelli theorem. 

     Next, from Theorem~\ref{thm:convergence}, $\widehat{\bm{w}}^{\mathrm{MMSCM}}_{T_0, G} \xrightarrow{\mathrm{p}} \bm{w}^*_G \in \widetilde{\Phi}^\dagger$ holds as $T_0 \to \infty$. 
     
     Lastly, $F_{Y^N_{j,t}, T_0}(y)$ converges uniformly in probability to $F_{Y^N_{0,t}}(y)$ as $T_0 \to \infty$ \mkato{under Assumptions~\ref{asm:bounded} and \ref{asm:mixture2} \citep{Potscher1989auniform}}. 

     By combining above results, we have
     \begin{align*}
        \sup_{y}\big|G_{T_0, L}(y) - F_{Y^N_{0,t}}(y)\big| \xrightarrow{\mathrm{p}} 0 \quad (L\to\infty, n\to\infty),
    \end{align*}
\end{proof}

\section{Confidence bands}
\label{appdx:confband}
\begin{figure}[t]
    \centering
            \includegraphics[width=0.95\linewidth]{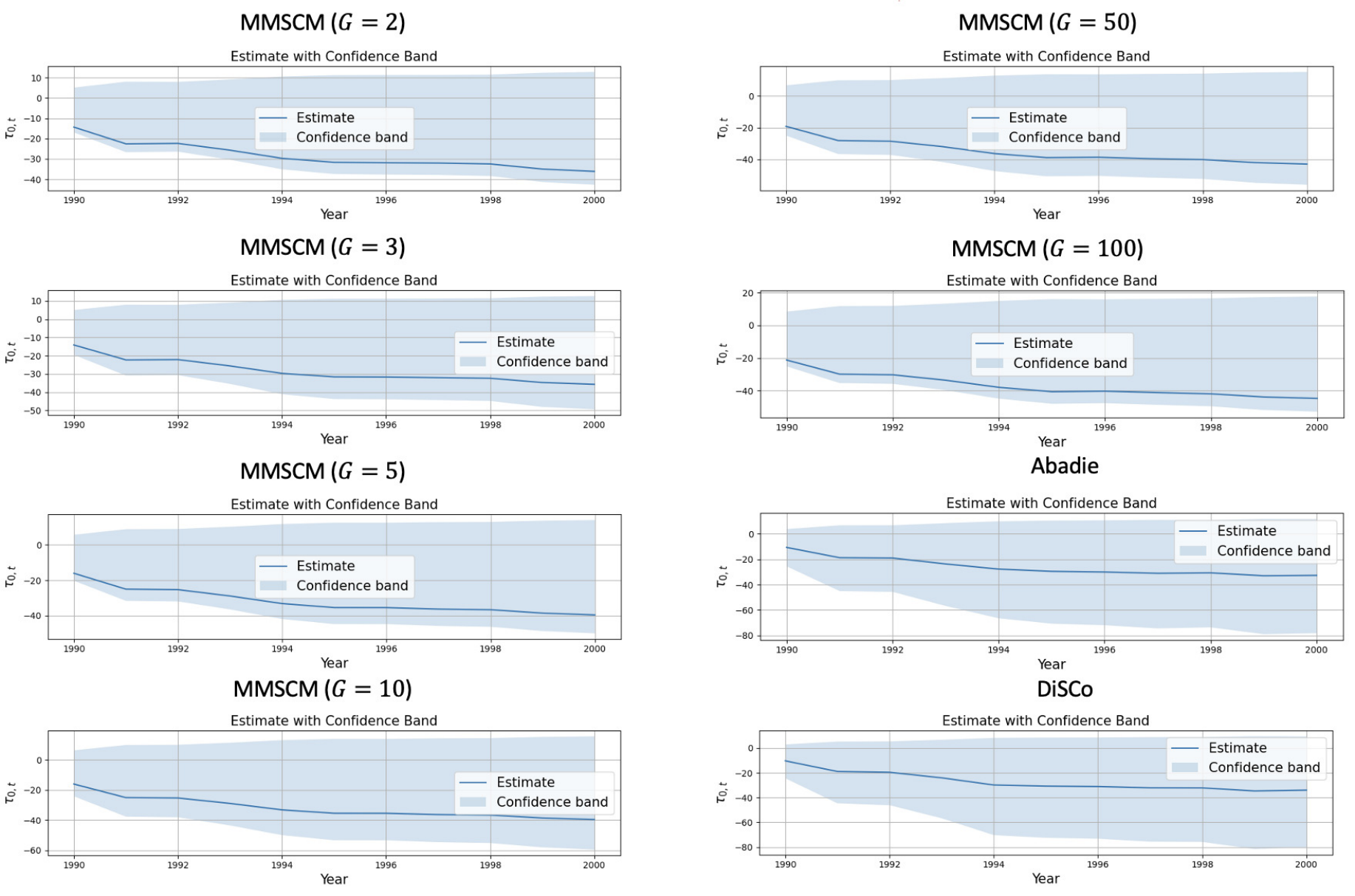}
        \caption{Confidence band given by conformal prediction with the ``Tobacco control program in California'' dataset.}
\label{fig:confinterval_tabacco}
\end{figure}

\begin{figure}[t]
    \centering
            \includegraphics[width=0.95\linewidth]{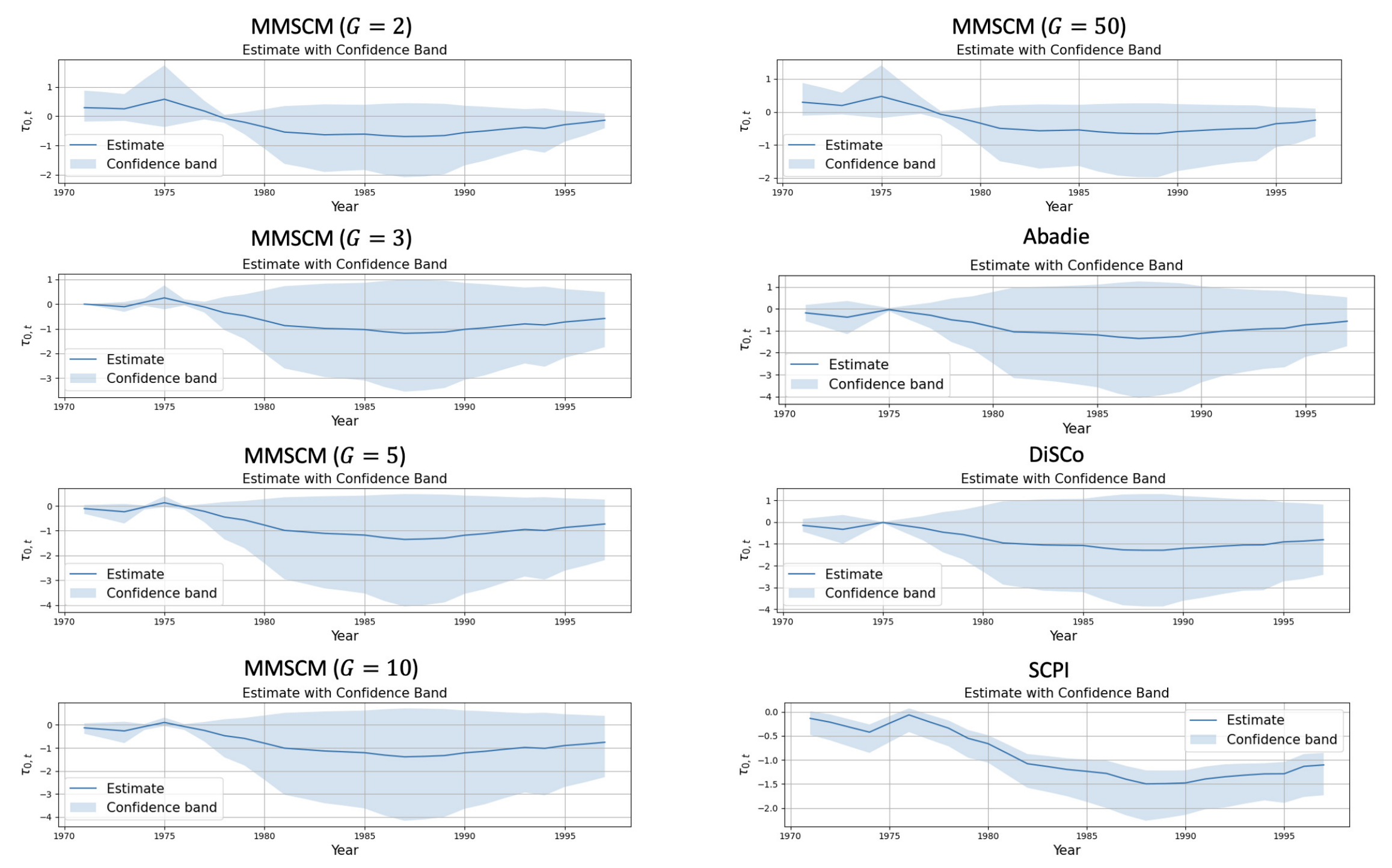}
        \caption{Confidence band given by conformal prediction with the ``Terrorist conflict in the Basque Country'' dataset.}
\label{fig:confinterval_basque}
\end{figure}

\begin{figure}[t]
    \centering
            \includegraphics[width=0.95\linewidth]{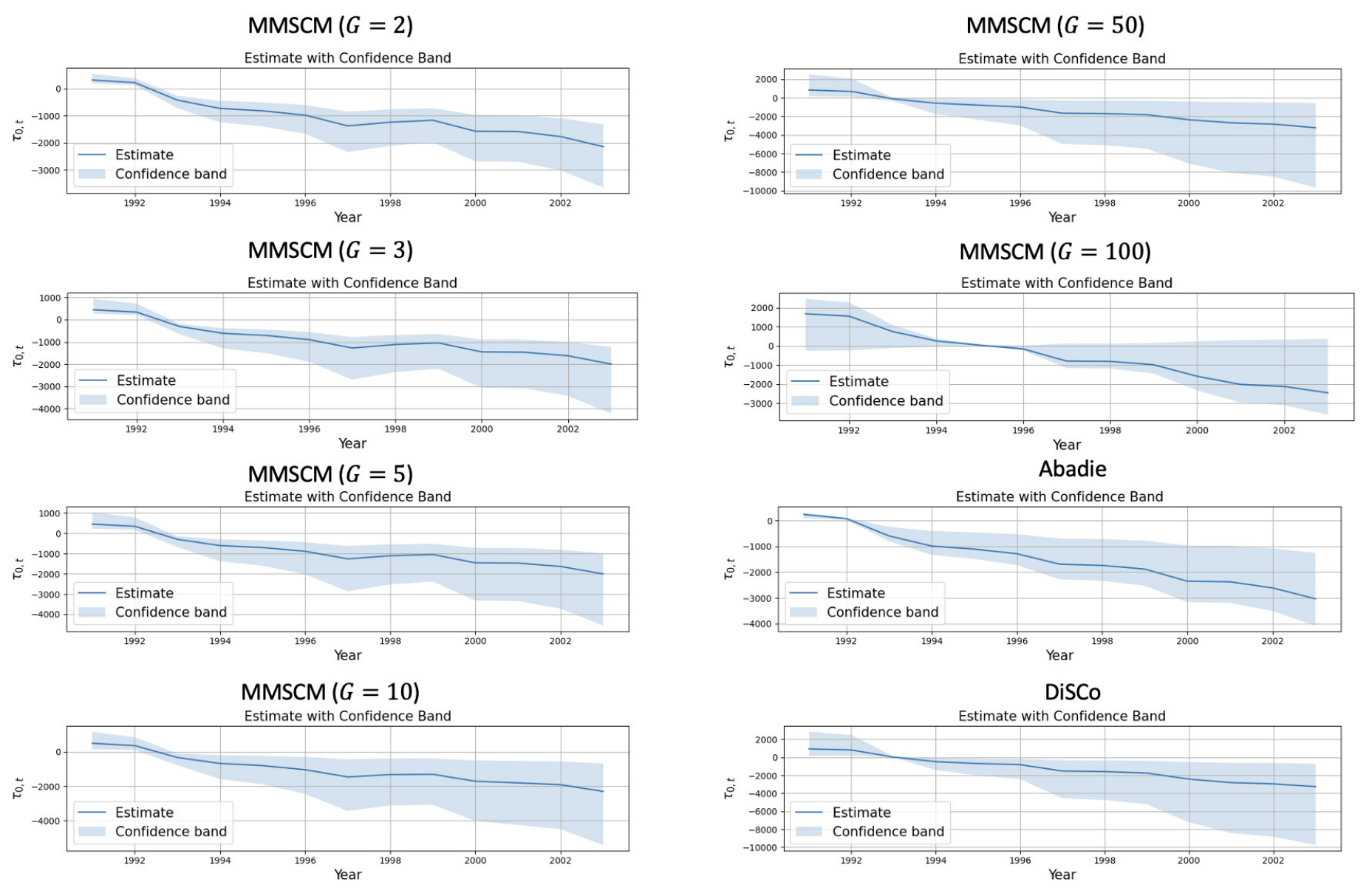}
        \caption{Confidence band given by conformal prediction with the ``Reunification of Germany'' dataset.}
\label{fig:confinterval_germany}
\end{figure}

In Tables~\ref{tab:conformal_tabacco}--\ref{tab:conformal_reunification} and Figures~\ref{fig:confinterval_tabacco}--\ref{fig:confinterval_germany}, we present the results of the conformal inference analysis described in Section~\ref{sec:empanalysis}. We omitted MMSCM ($G = 100$) due to the space limitation.

\color{black}

\bibliographystyle{qe}
\bibliography{distscm} 

\begin{thebibliography}{38}
\providecommand{\natexlab}[1]{#1}
\providecommand{\url}[1]{\texttt{#1}}
\expandafter\ifx\csname urlstyle\endcsname\relax
  \providecommand{\doi}[1]{doi: #1}\else
  \providecommand{\doi}{doi: \begingroup \urlstyle{rm}\Url}\fi

\bibitem[Abadie(2002)]{Abadie2002}
Alberto Abadie.
\newblock Bootstrap tests for distributional treatment effects in instrumental
  variable models.
\newblock \emph{Journal of the American Statistical Association}, 97\penalty0
  (457):\penalty0 284--292, 2002.

\bibitem[Abadie \& Gardeazabal(2003)Abadie and
  Gardeazabal]{AbadieGardeazabal2003}
Alberto Abadie and Javier Gardeazabal.
\newblock The economic costs of conflict: A case study of the basque country.
\newblock \emph{American Economic Review}, 93\penalty0 (1):\penalty0 113--132,
  2003.

\bibitem[Abadie et~al.(2010)Abadie, Diamond, and
  Hainmueller]{AbadieDiamondHainmueller2010}
Alberto Abadie, Alexis Diamond, and Jens Hainmueller.
\newblock Synthetic control methods for comparative case studies: Estimating
  the effect of california’s tobacco control program.
\newblock \emph{Journal of the American Statistical Association}, 105\penalty0
  (490):\penalty0 493--505, 2010.

\bibitem[Abadie et~al.(2015)Abadie, Diamond, and
  Hainmueller]{AbadieDiamondHainmueller2015}
Alberto Abadie, Alexis Diamond, and Jens Hainmueller.
\newblock Comparative politics and the synthetic control method.
\newblock \emph{American Journal of Political Science}, 59\penalty0
  (2):\penalty0 495--510, 2015.

\bibitem[Ben-Michael et~al.(2019)Ben-Michael, Feller, and Rothstein]{Ben2019}
Eli Ben-Michael, Avi Feller, and Jesse Rothstein.
\newblock Synthetic controls with staggered adoption, 2019.

\bibitem[Ben-Michael et~al.(2021)Ben-Michael, Feller, and Rothstein]{Ben2021}
Eli Ben-Michael, Avi Feller, and Jesse Rothstein.
\newblock The augmented synthetic control method.
\newblock \emph{Journal of the American Statistical Association}, 116\penalty0
  (536):\penalty0 1789--1803, 2021.

\bibitem[Carrasco \& Kotchoni(2017)Carrasco and
  Kotchoni]{Carrasco2017efficientestimation}
Marine Carrasco and Rachidi Kotchoni.
\newblock Efficient gmm estimation using the empirical characteristic function.
\newblock \emph{Econometric Theory}, 33\penalty0 (2):\penalty0 479--526, 2017.
\newblock ISSN 02664666, 14694360.

\bibitem[Cattaneo et~al.(2021)Cattaneo, Feng, and Titiunik]{CattaneoFeng2021}
Matias~D. Cattaneo, Yingjie Feng, and Rocio Titiunik.
\newblock Prediction intervals for synthetic control methods.
\newblock \emph{Journal of the American Statistical Association}, 116\penalty0
  (536):\penalty0 1865--1880, 2021.

\bibitem[Cattaneo et~al.(2022)Cattaneo, Feng, Palomba, and
  Titiunik]{Cattaneo2022scpiuncertainty}
Matias~D. Cattaneo, Yingjie Feng, Filippo Palomba, and Rocio Titiunik.
\newblock scpi: Uncertainty quantification for synthetic control methods, 2022.
\newblock {a}rXiv: 2202.05984.

\bibitem[Chen(2020)]{Chen2020}
Yi-Ting Chen.
\newblock A distributional synthetic control method for policy evaluation.
\newblock \emph{Journal of Applied Econometrics}, 35\penalty0 (5):\penalty0
  505--525, 2020.

\bibitem[Chernozhukov et~al.(2021)Chernozhukov, Wüthrich, and
  Zhu]{Chernozhukov2021}
Victor Chernozhukov, Kaspar Wüthrich, and Yinchu Zhu.
\newblock An exact and robust conformal inference method for counterfactual and
  synthetic controls.
\newblock \emph{Journal of the American Statistical Association}, 116\penalty0
  (536):\penalty0 1849--1864, 2021.

\bibitem[Chikahara et~al.(2022)Chikahara, Yamada, and
  Kashima]{chikahara2022feature}
Yoichi Chikahara, Makoto Yamada, and Hisashi Kashima.
\newblock Feature selection for discovering distributional treatment effect
  modifiers.
\newblock In \emph{UAI}, 2022.

\bibitem[Cunningham \& Shah(2017)Cunningham and Shah]{Cunningham2017}
Scott Cunningham and Manisha Shah.
\newblock {Decriminalizing Indoor Prostitution: Implications for Sexual
  Violence and Public Health}.
\newblock \emph{The Review of Economic Studies}, 85\penalty0 (3):\penalty0
  1683--1715, 2017.

\bibitem[Doudchenko \& Imbens(2016)Doudchenko and Imbens]{Doudchenko2016}
Nikolay Doudchenko and Guido~W Imbens.
\newblock Balancing, regression, difference-in-differences and synthetic
  control methods: A synthesis.
\newblock Working Paper 22791, National Bureau of Economic Research, 2016.

\bibitem[Ferman(2021)]{Ferman2021OntheProp}
Bruno Ferman.
\newblock On the properties of the synthetic control estimator with many
  periods and many controls.
\newblock \emph{Journal of the American Statistical Association}, 116\penalty0
  (536):\penalty0 1764--1772, 2021.

\bibitem[Ferman \& Pinto(2021)Ferman and Pinto]{Ferman2021}
Bruno Ferman and Cristine Pinto.
\newblock Synthetic controls with imperfect pretreatment fit.
\newblock \emph{Quantitative Economics}, 12\penalty0 (4):\penalty0 1197--1221,
  2021.

\bibitem[Fry(2024)]{fry2024method}
Joseph Fry.
\newblock A method of moments approach to asymptotically unbiased synthetic
  controls, 2024.
\newblock {a}rXiv:2312.01209.

\bibitem[Greene(2003)]{Greene2003Econometric}
William~H. Greene.
\newblock \emph{Econometric Analysis}.
\newblock Pearson Education, fifth edition, 2003.

\bibitem[Gretton et~al.(2012)Gretton, Borgwardt, Rasch, Sch{{\"o}}lkopf, and
  Smola]{Gretton2012}
Arthur Gretton, Karsten~M. Borgwardt, Malte~J. Rasch, Bernhard Sch{{\"o}}lkopf,
  and Alexander Smola.
\newblock A kernel two-sample test.
\newblock \emph{Journal of Machine Learning Research}, 13\penalty0
  (25):\penalty0 723--773, 2012.

\bibitem[Gunsilius(2023)]{Gunsilius2020}
F.~F. Gunsilius.
\newblock Distributional synthetic controls.
\newblock \emph{Econometrica}, 91\penalty0 (3):\penalty0 1105--1117, 2023.

\bibitem[Hausdorff(1921)]{Hausdorff1921summationsmethodenund}
Felix Hausdorff.
\newblock Summationsmethoden und momentfolgen. ii.
\newblock \emph{Mathematische Zeitschrift}, 9\penalty0 (3):\penalty0 280--299,
  Sep 1921.
\newblock ISSN 1432-1823.
\newblock \doi{10.1007/BF01279032}.
\newblock URL \url{https://doi.org/10.1007/BF01279032}.

\bibitem[Kallus \& Oprescu(2022)Kallus and Oprescu]{KallusOprescu2022}
Nathan Kallus and Miruna Oprescu.
\newblock Robust and agnostic learning of conditional distributional treatment
  effects, 2022.
\newblock {a}rXiv:2205.11486.

\bibitem[Kato \& Teshima(2021)Kato and Teshima]{Kato2021bregman}
Masahiro Kato and Takeshi Teshima.
\newblock Non-negative bregman divergence minimization for deep direct density
  ratio estimation.
\newblock In \emph{International Conference on Machine Learning}, volume 139,
  pp.\  5320--5333, 2021.

\bibitem[Li(2020)]{Kathleen2020}
Kathleen~T. Li.
\newblock Statistical inference for average treatment effects estimated by
  synthetic control methods.
\newblock \emph{Journal of the American Statistical Association}, 115\penalty0
  (532):\penalty0 2068--2083, 2020.

\bibitem[Maier(2011)]{Maier2011}
Michael Maier.
\newblock Tests for distributional treatment effects under unconfoundedness.
\newblock \emph{Economics Letters}, 110\penalty0 (1):\penalty0 49--51, 2011.

\bibitem[Nazaret et~al.(2023)Nazaret, Shi, and
  Blei]{nazaret2023misspecification}
Achille Nazaret, Claudia Shi, and David~M. Blei.
\newblock On the misspecification of linear assumptions in synthetic control,
  2023.

\bibitem[Neyman(1923)]{Neyman1923}
Jerzy Neyman.
\newblock Sur les applications de la theorie des probabilites aux experiences
  agricoles: Essai des principes.
\newblock \emph{Statistical Science}, 5:\penalty0 463–472, 1923.

\bibitem[Park et~al.(2021)Park, Shalit, Sch{\"o}lkopf, and
  Muandet]{ParkShalit2021}
Junhyung Park, Uri Shalit, Bernhard Sch{\"o}lkopf, and Krikamol Muandet.
\newblock Conditional distributional treatment effect with kernel conditional
  mean embeddings and u-statistic regression.
\newblock In \emph{ICML}, pp.\  8401--8412, 2021.

\bibitem[Pötscher \& Prucha(1989)Pötscher and Prucha]{Potscher1989auniform}
Benedikt~M. Pötscher and Ingmar~R. Prucha.
\newblock A uniform law of large numbers for dependent and heterogeneous data
  processes.
\newblock \emph{Econometrica}, 57\penalty0 (3):\penalty0 675--683, 1989.

\bibitem[Ritzwoller et~al.(2025)Ritzwoller, Romano, and
  Shaikh]{Ritzwoller2025randomizationinference}
David~M. Ritzwoller, Joseph~P. Romano, and Azeem~M. Shaikh.
\newblock Randomization inference: Theory and applications, 2025.
\newblock {a}rXiv: 2406.09521.

\bibitem[Rubin(1974)]{Rubin1974}
Donald~B. Rubin.
\newblock Estimating causal effects of treatments in randomized and
  nonrandomized studies.
\newblock \emph{Journal of Educational Psychology}, 1974.

\bibitem[Schmüdgen(2020)]{Stchmudgen2020lecturesmomentproblem}
Konrad Schmüdgen.
\newblock Ten lectures on the moment problem, 2020.
\newblock {a}rXiv: 2008.12698.

\bibitem[Shaikh \& Toulis(2021)Shaikh and Toulis]{Shaikh2021}
Azeem~M. Shaikh and Panos Toulis.
\newblock Randomization tests in observational studies with staggered adoption
  of treatment.
\newblock \emph{Journal of the American Statistical Association}, 116\penalty0
  (536):\penalty0 1835--1848, 2021.

\bibitem[Shi et~al.(2022)Shi, Sridhar, Misra, and Blei]{Shi2022}
Claudia Shi, Dhanya Sridhar, Vishal Misra, and David Blei.
\newblock On the assumptions of synthetic control methods.
\newblock In \emph{AISTATS}, pp.\  7163--7175, 2022.

\bibitem[Spiess et~al.(2023)Spiess, Imbens, and Venugopal]{spiess2023double}
Jann Spiess, Guido Imbens, and Amar Venugopal.
\newblock Double and single descent in causal inference with an application to
  high-dimensional synthetic control, 2023.

\bibitem[Sugiyama et~al.(2012)Sugiyama, Suzuki, and
  Kanamori]{Sugiyama:2012:DRE:2181148}
Masashi Sugiyama, Taiji Suzuki, and Takafumi Kanamori.
\newblock \emph{Density Ratio Estimation in Machine Learning}.
\newblock 2012.

\bibitem[Vovk et~al.(2005)Vovk, Gammerman, and Shafer]{Vovk2005}
Vladimir Vovk, Alex Gammerman, and Glenn Shafer.
\newblock \emph{Algorithmic Learning in a Random World}.
\newblock Springer-Verlag, 2005.

\bibitem[Wan et~al.(2018)Wan, Xie, and Hsiao]{Wan2018}
Shui-Ki Wan, Yimeng Xie, and Cheng Hsiao.
\newblock Panel data approach vs synthetic control method.
\newblock \emph{Economics Letters}, 164:\penalty0 121--123, 2018.
\newblock \doi{https://doi.org/10.1016/j.econlet.2018.01.019}.

\end{thebibliography}

\clearpage

\begin{table}[t]
\caption{Results of treatment effect estimation and conformal inference with the ``Tobacco control program in California'' dataset. Estimated treatment effect $\widehat{\tau}_{0, t}$ and $90\%$ confidence interval $(\underline{\tau}_{0, t}, \overline{\tau}_{0, t})$.}
\label{tab:conformal_tabacco}
\centering
\caption*{MMSCM ($G = 2$)}
\scalebox{0.35}{ 
\begin{tabular}{lrrrrrrrrrrrrrrrrrrrrrrrrrrr}
\toprule
 & 1971 & 1972 & 1973 & 1974 & 1975 & 1976 & 1977 & 1978 & 1979 & 1980 & 1981 & 1982 & 1983 & 1984 & 1985 & 1986 & 1987 & 1988 & 1989 & 1990 & 1991 & 1992 & 1993 & 1994 & 1995 & 1996 & 1997 \\
\midrule
Estimate ($\widehat{\tau}_{0, t}$) & 0.290 & 0.274 & 0.251 & 0.421 & 0.578 & 0.370 & 0.175 & -0.076 & -0.206 & -0.369 & -0.543 & -0.585 & -0.635 & -0.621 & -0.612 & -0.664 & -0.694 & -0.685 & -0.660 & -0.559 & -0.506 & -0.439 & -0.379 & -0.415 & -0.289 & -0.220 & -0.137 \\
Lower bound ($\underline{\tau}_{0, t}$) & -0.184 & -0.174 & -0.160 & -0.268 & -0.368 & -0.236 & -0.111 & -0.228 & -0.618 & -1.107 & -1.629 & -1.755 & -1.904 & -1.863 & -1.835 & -1.992 & -2.083 & -2.055 & -1.980 & -1.678 & -1.519 & -1.318 & -1.137 & -1.245 & -0.868 & -0.660 & -0.411 \\
Upper bound ($\overline{\tau}_{0, t}$) & 0.870 & 0.822 & 0.753 & 1.262 & 1.734 & 1.111 & 0.524 & 0.048 & 0.131 & 0.235 & 0.346 & 0.372 & 0.404 & 0.395 & 0.389 & 0.423 & 0.442 & 0.436 & 0.420 & 0.356 & 0.322 & 0.279 & 0.241 & 0.264 & 0.184 & 0.140 & 0.087 \\
\bottomrule
\end{tabular}
}
\vspace{5mm}
\caption*{MMSCM ($G = 3$)}
\scalebox{0.35}{ 
\begin{tabular}{lrrrrrrrrrrrrrrrrrrrrrrrrrrr}
\toprule
 & 1971 & 1972 & 1973 & 1974 & 1975 & 1976 & 1977 & 1978 & 1979 & 1980 & 1981 & 1982 & 1983 & 1984 & 1985 & 1986 & 1987 & 1988 & 1989 & 1990 & 1991 & 1992 & 1993 & 1994 & 1995 & 1996 & 1997 \\
\midrule
Estimate ($\widehat{\tau}_{0, t}$) & -0.001 & -0.051 & -0.105 & 0.079 & 0.253 & 0.063 & -0.115 & -0.351 & -0.473 & -0.668 & -0.872 & -0.927 & -0.988 & -1.008 & -1.034 & -1.123 & -1.185 & -1.168 & -1.136 & -1.025 & -0.965 & -0.881 & -0.804 & -0.847 & -0.726 & -0.658 & -0.583 \\
Lower bound ($\underline{\tau}_{0, t}$) & -0.004 & -0.153 & -0.315 & -0.066 & -0.212 & -0.053 & -0.344 & -1.053 & -1.418 & -2.003 & -2.615 & -2.781 & -2.965 & -3.025 & -3.101 & -3.369 & -3.554 & -3.505 & -3.408 & -3.076 & -2.896 & -2.643 & -2.413 & -2.542 & -2.178 & -1.975 & -1.748 \\
Upper bound ($\overline{\tau}_{0, t}$) & 0.001 & 0.043 & 0.088 & 0.237 & 0.759 & 0.190 & 0.096 & 0.294 & 0.396 & 0.560 & 0.731 & 0.777 & 0.829 & 0.845 & 0.867 & 0.942 & 0.993 & 0.980 & 0.952 & 0.860 & 0.809 & 0.739 & 0.674 & 0.710 & 0.609 & 0.552 & 0.488 \\
\bottomrule
\end{tabular}
}
\vspace{5mm}
\caption*{MMSCM ($G = 5$)}
\scalebox{0.35}{ 
\begin{tabular}{lrrrrrrrrrrrrrrrrrrrrrrrrrrr}
\toprule
 & 1971 & 1972 & 1973 & 1974 & 1975 & 1976 & 1977 & 1978 & 1979 & 1980 & 1981 & 1982 & 1983 & 1984 & 1985 & 1986 & 1987 & 1988 & 1989 & 1990 & 1991 & 1992 & 1993 & 1994 & 1995 & 1996 & 1997 \\
\midrule
Estimate ($\widehat{\tau}_{0, t}$) & -0.129 & -0.195 & -0.263 & -0.074 & 0.105 & -0.074 & -0.242 & -0.471 & -0.588 & -0.796 & -1.012 & -1.071 & -1.135 & -1.170 & -1.209 & -1.314 & -1.388 & -1.368 & -1.332 & -1.215 & -1.150 & -1.060 & -0.980 & -1.020 & -0.898 & -0.832 & -0.757 \\
Lower bound ($\underline{\tau}_{0, t}$) & -0.388 & -0.585 & -0.790 & -0.223 & -0.054 & -0.222 & -0.727 & -1.412 & -1.763 & -2.388 & -3.036 & -3.212 & -3.404 & -3.509 & -3.628 & -3.943 & -4.165 & -4.105 & -3.996 & -3.644 & -3.450 & -3.181 & -2.939 & -3.061 & -2.695 & -2.495 & -2.272 \\
Upper bound ($\overline{\tau}_{0, t}$) & 0.067 & 0.100 & 0.136 & 0.038 & 0.316 & 0.038 & 0.125 & 0.242 & 0.303 & 0.410 & 0.521 & 0.552 & 0.585 & 0.602 & 0.623 & 0.677 & 0.715 & 0.705 & 0.686 & 0.626 & 0.592 & 0.546 & 0.505 & 0.526 & 0.463 & 0.428 & 0.390 \\
\bottomrule
\end{tabular}
}
\vspace{5mm}
\caption*{MMSCM ($G = 10$)}
\scalebox{0.35}{ 
\begin{tabular}{lrrrrrrrrrrrrrrrrrrrrrrrrrrr}
\toprule
 & 1971 & 1972 & 1973 & 1974 & 1975 & 1976 & 1977 & 1978 & 1979 & 1980 & 1981 & 1982 & 1983 & 1984 & 1985 & 1986 & 1987 & 1988 & 1989 & 1990 & 1991 & 1992 & 1993 & 1994 & 1995 & 1996 & 1997 \\
\midrule
Estimate ($\widehat{\tau}_{0, t}$) & -0.107 & -0.170 & -0.236 & -0.048 & 0.130 & -0.051 & -0.220 & -0.450 & -0.568 & -0.774 & -0.988 & -1.046 & -1.110 & -1.141 & -1.178 & -1.281 & -1.353 & -1.334 & -1.299 & -1.183 & -1.120 & -1.031 & -0.951 & -0.992 & -0.871 & -0.804 & -0.730 \\
Lower bound ($\underline{\tau}_{0, t}$) & -0.322 & -0.511 & -0.709 & -0.145 & -0.046 & -0.152 & -0.661 & -1.350 & -1.704 & -2.321 & -2.963 & -3.138 & -3.329 & -3.424 & -3.535 & -3.842 & -4.058 & -4.002 & -3.896 & -3.550 & -3.359 & -3.093 & -2.853 & -2.977 & -2.612 & -2.412 & -2.190 \\
Upper bound ($\overline{\tau}_{0, t}$) & 0.038 & 0.060 & 0.084 & 0.017 & 0.391 & 0.018 & 0.078 & 0.159 & 0.201 & 0.274 & 0.349 & 0.370 & 0.392 & 0.403 & 0.417 & 0.453 & 0.478 & 0.472 & 0.459 & 0.418 & 0.396 & 0.365 & 0.336 & 0.351 & 0.308 & 0.284 & 0.258 \\
\bottomrule
\end{tabular}
}
\vspace{5mm}
\caption*{MMSCM ($G = 50$)}
\scalebox{0.35}{ 
\begin{tabular}{lrrrrrrrrrrrrrrrrrrrrrrrrrrr}
\toprule
 & 1971 & 1972 & 1973 & 1974 & 1975 & 1976 & 1977 & 1978 & 1979 & 1980 & 1981 & 1982 & 1983 & 1984 & 1985 & 1986 & 1987 & 1988 & 1989 & 1990 & 1991 & 1992 & 1993 & 1994 & 1995 & 1996 & 1997 \\
\midrule
Estimate ($\widehat{\tau}_{0, t}$) & 0.293 & 0.246 & 0.194 & 0.337 & 0.471 & 0.309 & 0.151 & -0.072 & -0.191 & -0.341 & -0.499 & -0.534 & -0.573 & -0.561 & -0.547 & -0.604 & -0.644 & -0.660 & -0.661 & -0.599 & -0.569 & -0.536 & -0.512 & -0.497 & -0.357 & -0.322 & -0.249 \\
Lower bound ($\underline{\tau}_{0, t}$) & -0.115 & -0.097 & -0.077 & -0.133 & -0.186 & -0.122 & -0.060 & -0.215 & -0.574 & -1.023 & -1.497 & -1.603 & -1.719 & -1.682 & -1.641 & -1.811 & -1.933 & -1.979 & -1.983 & -1.797 & -1.706 & -1.609 & -1.535 & -1.491 & -1.070 & -0.967 & -0.747 \\
Upper bound ($\overline{\tau}_{0, t}$) & 0.879 & 0.738 & 0.583 & 1.012 & 1.413 & 0.926 & 0.453 & 0.028 & 0.075 & 0.134 & 0.197 & 0.211 & 0.226 & 0.221 & 0.215 & 0.238 & 0.254 & 0.260 & 0.260 & 0.236 & 0.224 & 0.211 & 0.202 & 0.196 & 0.141 & 0.127 & 0.098 \\
\bottomrule
\end{tabular}
}
\vspace{5mm}
\caption*{MMSCM ($G = 100$)}
\scalebox{0.35}{ 
\begin{tabular}{lrrrrrrrrrrrrrrrrrrrrrrrrrrr}
\toprule
 & 1971 & 1972 & 1973 & 1974 & 1975 & 1976 & 1977 & 1978 & 1979 & 1980 & 1981 & 1982 & 1983 & 1984 & 1985 & 1986 & 1987 & 1988 & 1989 & 1990 & 1991 & 1992 & 1993 & 1994 & 1995 & 1996 & 1997 \\
\midrule
Estimate ($\widehat{\tau}_{0, t}$) & 0.381 & 0.338 & 0.290 & 0.420 & 0.539 & 0.386 & 0.236 & 0.021 & -0.098 & -0.231 & -0.373 & -0.396 & -0.422 & -0.391 & -0.356 & -0.400 & -0.433 & -0.458 & -0.470 & -0.416 & -0.389 & -0.376 & -0.369 & -0.335 & -0.186 & -0.158 & -0.083 \\
Lower bound ($\underline{\tau}_{0, t}$) & -0.288 & -0.256 & -0.220 & -0.318 & -0.409 & -0.292 & -0.179 & -0.016 & -0.293 & -0.693 & -1.120 & -1.189 & -1.266 & -1.172 & -1.068 & -1.201 & -1.298 & -1.374 & -1.410 & -1.249 & -1.168 & -1.127 & -1.107 & -1.005 & -0.557 & -0.473 & -0.250 \\
Upper bound ($\overline{\tau}_{0, t}$) & 1.142 & 1.013 & 0.871 & 1.259 & 1.618 & 1.158 & 0.707 & 0.064 & 0.074 & 0.175 & 0.283 & 0.300 & 0.320 & 0.296 & 0.270 & 0.303 & 0.328 & 0.347 & 0.356 & 0.315 & 0.295 & 0.285 & 0.280 & 0.254 & 0.141 & 0.119 & 0.063 \\
\bottomrule
\end{tabular}
}
\vspace{5mm}
\caption*{Abadie}
\scalebox{0.35}{ 
\begin{tabular}{lrrrrrrrrrrrrrrrrrrrrrrrrrrr}
\toprule
 & 1971 & 1972 & 1973 & 1974 & 1975 & 1976 & 1977 & 1978 & 1979 & 1980 & 1981 & 1982 & 1983 & 1984 & 1985 & 1986 & 1987 & 1988 & 1989 & 1990 & 1991 & 1992 & 1993 & 1994 & 1995 & 1996 & 1997 \\
\midrule
Estimate ($\widehat{\tau}_{0, t}$) & -0.188 & -0.286 & -0.385 & -0.208 & -0.038 & -0.172 & -0.297 & -0.505 & -0.615 & -0.831 & -1.052 & -1.077 & -1.105 & -1.147 & -1.192 & -1.289 & -1.354 & -1.316 & -1.263 & -1.119 & -1.023 & -0.964 & -0.913 & -0.889 & -0.731 & -0.662 & -0.569 \\
Lower bound ($\underline{\tau}_{0, t}$) & -0.565 & -0.859 & -1.155 & -0.624 & -0.113 & -0.515 & -0.892 & -1.516 & -1.845 & -2.492 & -3.157 & -3.230 & -3.316 & -3.441 & -3.576 & -3.866 & -4.063 & -3.949 & -3.788 & -3.356 & -3.069 & -2.891 & -2.739 & -2.667 & -2.192 & -1.985 & -1.707 \\
Upper bound ($\overline{\tau}_{0, t}$) & 0.173 & 0.263 & 0.354 & 0.191 & 0.035 & 0.158 & 0.273 & 0.465 & 0.565 & 0.764 & 0.967 & 0.990 & 1.016 & 1.054 & 1.096 & 1.184 & 1.245 & 1.210 & 1.161 & 1.028 & 0.940 & 0.886 & 0.839 & 0.817 & 0.672 & 0.608 & 0.523 \\
\bottomrule
\end{tabular}
}
\vspace{5mm}
\caption*{DiSCo}
\scalebox{0.35}{ 
\begin{tabular}{lrrrrrrrrrrrrrrrrrrrrrrrrrrr}
\toprule
 & 1971 & 1972 & 1973 & 1974 & 1975 & 1976 & 1977 & 1978 & 1979 & 1980 & 1981 & 1982 & 1983 & 1984 & 1985 & 1986 & 1987 & 1988 & 1989 & 1990 & 1991 & 1992 & 1993 & 1994 & 1995 & 1996 & 1997 \\
\midrule
Estimate ($\widehat{\tau}_{0, t}$) & -0.147 & -0.238 & -0.330 & -0.169 & -0.015 & -0.149 & -0.278 & -0.463 & -0.569 & -0.758 & -0.957 & -1.002 & -1.048 & -1.061 & -1.072 & -1.185 & -1.273 & -1.290 & -1.290 & -1.204 & -1.155 & -1.098 & -1.050 & -1.044 & -0.907 & -0.870 & -0.807 \\
Lower bound ($\underline{\tau}_{0, t}$) & -0.440 & -0.715 & -0.990 & -0.508 & -0.046 & -0.448 & -0.834 & -1.388 & -1.707 & -2.274 & -2.870 & -3.006 & -3.145 & -3.184 & -3.215 & -3.555 & -3.820 & -3.869 & -3.871 & -3.611 & -3.466 & -3.295 & -3.149 & -3.131 & -2.720 & -2.609 & -2.420 \\
Upper bound ($\overline{\tau}_{0, t}$) & 0.147 & 0.238 & 0.330 & 0.169 & 0.015 & 0.149 & 0.278 & 0.463 & 0.569 & 0.758 & 0.957 & 1.002 & 1.048 & 1.061 & 1.072 & 1.185 & 1.273 & 1.290 & 1.290 & 1.204 & 1.155 & 1.098 & 1.050 & 1.044 & 0.907 & 0.870 & 0.807 \\
\bottomrule
\end{tabular}
}
\vspace{5mm}
\caption*{SCPI}
\scalebox{0.35}{ 
\begin{tabular}{lrrrrrrrrrrrrrrrrrrrrrrrrrrr}
\toprule
 & 1971 & 1972 & 1973 & 1974 & 1975 & 1976 & 1977 & 1978 & 1979 & 1980 & 1981 & 1982 & 1983 & 1984 & 1985 & 1986 & 1987 & 1988 & 1989 & 1990 & 1991 & 1992 & 1993 & 1994 & 1995 & 1996 & 1997 \\
\midrule
Estimate ($\widehat{\tau}_{0, t}$) & -0.138 & -0.216 & -0.321 & -0.424 & -0.241 & -0.065 & -0.206 & -0.340 & -0.553 & -0.662 & -0.869 & -1.080 & -1.137 & -1.199 & -1.238 & -1.281 & -1.403 & -1.498 & -1.493 & -1.479 & -1.397 & -1.351 & -1.317 & -1.291 & -1.288 & -1.134 & -1.105 \\
Lower bound ($\underline{\tau}_{0, t}$) & 0.011 & -0.055 & -0.160 & -0.264 & -0.094 & 0.069 & -0.058 & -0.181 & -0.379 & -0.486 & -0.680 & -0.879 & -0.921 & -0.968 & -0.990 & -1.014 & -1.128 & -1.218 & -1.222 & -1.217 & -1.139 & -1.092 & -1.077 & -1.070 & -1.044 & -0.879 & -0.856 \\
Upper bound ($\overline{\tau}_{0, t}$) & -0.481 & -0.593 & -0.725 & -0.855 & -0.636 & -0.422 & -0.574 & -0.716 & -0.955 & -1.056 & -1.319 & -1.580 & -1.664 & -1.754 & -1.872 & -2.001 & -2.152 & -2.264 & -2.208 & -2.141 & -2.026 & -1.989 & -1.909 & -1.841 & -1.892 & -1.769 & -1.735 \\
\bottomrule
\end{tabular}

}
\end{table}

\begin{table}[t]
\caption{Results of treatment effect estimation and conformal inference with the ``Terrorist conflict in the Basque Country'' dataset. Estimated treatment effect $\widehat{\tau}_{0, t}$ and $90\%$ confidence interval $(\underline{\tau}_{0, t}, \overline{\tau}_{0, t})$.}
\label{tab:conformal_terrorist}
\centering
\vspace{3.5mm}
\caption*{MMSCM ($G = 2$)}
\vspace{-2mm}
\scalebox{0.50}{ 
\begin{tabular}{lrrrrrrrrrrrrr}
\toprule
 & 1991 & 1992 & 1993 & 1994 & 1995 & 1996 & 1997 & 1998 & 1999 & 2000 & 2001 & 2002 & 2003 \\
\midrule
Estimate ($\widehat{\tau}_{0, t}$) & 327.880 & 228.558 & -418.809 & -724.432 & -815.110 & -974.383 & -1372.114 & -1235.992 & -1161.941 & -1571.170 & -1579.920 & -1769.058 & -2140.223 \\
Lower bound ($\underline{\tau}_{0, t}$) & 202.027 & 140.829 & -714.936 & -1236.657 & -1391.451 & -1663.340 & -2342.296 & -2109.926 & -1983.516 & -2682.097 & -2697.035 & -3019.907 & -3653.512 \\
Upper bound ($\overline{\tau}_{0, t}$) & 559.715 & 390.165 & -258.054 & -446.367 & -502.240 & -600.377 & -845.444 & -761.571 & -715.944 & -968.094 & -973.486 & -1090.026 & -1318.723 \\
\bottomrule
\end{tabular}
}
\vspace{3.5mm}
\caption*{MMSCM ($G = 3$)}
\vspace{-2mm}
\scalebox{0.50}{ 
\begin{tabular}{lrrrrrrrrrrrrr}
\toprule
 & 1991 & 1992 & 1993 & 1994 & 1995 & 1996 & 1997 & 1998 & 1999 & 2000 & 2001 & 2002 & 2003 \\
\midrule
Estimate ($\widehat{\tau}_{0, t}$) & 445.970 & 341.007 & -307.692 & -612.796 & -710.657 & -896.374 & -1274.747 & -1114.977 & -1045.510 & -1450.164 & -1463.503 & -1624.141 & -1995.589 \\
Lower bound ($\underline{\tau}_{0, t}$) & 274.790 & 210.115 & -649.573 & -1293.680 & -1500.275 & -1892.345 & -2691.132 & -2353.840 & -2207.188 & -3061.457 & -3089.617 & -3428.742 & -4212.909 \\
Upper bound ($\overline{\tau}_{0, t}$) & 941.492 & 719.904 & -189.588 & -377.581 & -437.879 & -552.311 & -785.450 & -687.006 & -644.203 & -893.535 & -901.754 & -1000.733 & -1229.605 \\
\bottomrule
\end{tabular}
}
\vspace{3.5mm}
\caption*{MMSCM ($G = 5$)}
\vspace{-2mm}
\scalebox{0.50}{ 
\begin{tabular}{lrrrrrrrrrrrrr}
\toprule
 & 1991 & 1992 & 1993 & 1994 & 1995 & 1996 & 1997 & 1998 & 1999 & 2000 & 2001 & 2002 & 2003 \\
\midrule
Estimate ($\widehat{\tau}_{0, t}$) & 454.533 & 347.786 & -297.835 & -605.164 & -703.606 & -888.683 & -1261.814 & -1109.264 & -1049.019 & -1458.731 & -1470.792 & -1635.784 & -2006.213 \\
Lower bound ($\underline{\tau}_{0, t}$) & 224.971 & 172.137 & -676.898 & -1375.372 & -1599.105 & -2019.734 & -2867.760 & -2521.054 & -2384.135 & -3315.299 & -3342.709 & -3717.691 & -4559.576 \\
Upper bound ($\overline{\tau}_{0, t}$) & 1033.028 & 790.423 & -147.413 & -299.525 & -348.249 & -439.853 & -624.534 & -549.029 & -519.212 & -721.998 & -727.968 & -809.631 & -992.974 \\
\bottomrule
\end{tabular}
}
\vspace{3.5mm}
\caption*{MMSCM ($G = 10$)}
\vspace{-2mm}
\scalebox{0.50}{ 
\begin{tabular}{lrrrrrrrrrrrrr}
\toprule
 & 1991 & 1992 & 1993 & 1994 & 1995 & 1996 & 1997 & 1998 & 1999 & 2000 & 2001 & 2002 & 2003 \\
\midrule
Estimate ($\widehat{\tau}_{0, t}$) & 496.735 & 362.964 & -330.006 & -673.647 & -801.127 & -1041.570 & -1461.328 & -1327.563 & -1308.178 & -1710.521 & -1808.892 & -1910.430 & -2305.259 \\
Lower bound ($\underline{\tau}_{0, t}$) & 145.508 & 106.323 & -776.680 & -1585.451 & -1885.482 & -2451.372 & -3439.287 & -3124.467 & -3078.843 & -4025.771 & -4257.291 & -4496.265 & -5425.507 \\
Upper bound ($\overline{\tau}_{0, t}$) & 1169.083 & 854.249 & -96.668 & -197.331 & -234.674 & -305.106 & -428.066 & -388.882 & -383.204 & -501.062 & -529.877 & -559.621 & -675.278 \\
\bottomrule
\end{tabular}
}
\vspace{3.5mm}
\caption*{MMSCM ($G = 50$)}
\vspace{-2mm}
\scalebox{0.50}{ 
\begin{tabular}{lrrrrrrrrrrrrr}
\toprule
 & 1991 & 1992 & 1993 & 1994 & 1995 & 1996 & 1997 & 1998 & 1999 & 2000 & 2001 & 2002 & 2003 \\
\midrule
Estimate ($\widehat{\tau}_{0, t}$) & 841.486 & 691.160 & -121.062 & -579.295 & -781.670 & -991.560 & -1654.087 & -1699.780 & -1822.579 & -2358.659 & -2690.635 & -2827.447 & -3223.693 \\
Lower bound ($\underline{\tau}_{0, t}$) & 144.498 & 118.684 & -363.185 & -1737.885 & -2345.010 & -2974.680 & -4962.262 & -5099.340 & -5467.738 & -7075.977 & -8071.906 & -8482.342 & -9671.078 \\
Upper bound ($\overline{\tau}_{0, t}$) & 2524.457 & 2073.479 & -20.788 & -99.475 & -134.226 & -170.268 & -284.035 & -291.881 & -312.968 & -405.022 & -462.028 & -485.521 & -553.563 \\
\bottomrule
\end{tabular}
}
\vspace{3.5mm}
\caption*{MMSCM ($G = 100$)}
\vspace{-2mm}
\scalebox{0.50}{ 
\begin{tabular}{lrrrrrrrrrrrrr}
\toprule
 & 1991 & 1992 & 1993 & 1994 & 1995 & 1996 & 1997 & 1998 & 1999 & 2000 & 2001 & 2002 & 2003 \\
\midrule
Estimate ($\widehat{\tau}_{0, t}$) & 1676.113 & 1555.814 & 749.219 & 260.681 & 41.160 & -157.906 & -792.896 & -807.118 & -986.725 & -1589.689 & -2015.736 & -2121.088 & -2451.771 \\
Lower bound ($\underline{\tau}_{0, t}$) & -253.957 & -235.729 & -113.518 & -39.497 & -6.236 & -231.277 & -1161.312 & -1182.143 & -1445.203 & -2328.332 & -2952.340 & -3106.644 & -3590.978 \\
Upper bound ($\overline{\tau}_{0, t}$) & 2454.913 & 2278.717 & 1097.341 & 381.805 & 60.284 & 23.925 & 120.136 & 122.291 & 149.504 & 240.862 & 305.414 & 321.377 & 371.480 \\
\bottomrule
\end{tabular}
}
\vspace{3.5mm}
\caption*{Abadie}
\vspace{-2mm}
\scalebox{0.50}{ 
\begin{tabular}{lrrrrrrrrrrrrr}
\toprule
 & 1991 & 1992 & 1993 & 1994 & 1995 & 1996 & 1997 & 1998 & 1999 & 2000 & 2001 & 2002 & 2003 \\
\midrule
Estimate ($\widehat{\tau}_{0, t}$) & 233.434 & 71.136 & -598.967 & -990.495 & -1110.130 & -1289.025 & -1693.204 & -1740.466 & -1884.240 & -2354.122 & -2377.322 & -2614.882 & -3033.735 \\
Lower bound ($\underline{\tau}_{0, t}$) & 96.675 & 29.460 & -804.673 & -1330.665 & -1491.387 & -1731.720 & -2274.708 & -2338.202 & -2531.353 & -3162.608 & -3193.776 & -3512.923 & -4075.623 \\
Upper bound ($\overline{\tau}_{0, t}$) & 313.604 & 95.566 & -248.057 & -410.205 & -459.751 & -533.839 & -701.226 & -720.799 & -780.342 & -974.939 & -984.547 & -1082.931 & -1256.395 \\
\bottomrule
\end{tabular}
}
\vspace{3.5mm}
\caption*{DiSCo}
\vspace{-2mm}
\scalebox{0.50}{ 
\begin{tabular}{lrrrrrrrrrrrrr}
\toprule
 & 1991 & 1992 & 1993 & 1994 & 1995 & 1996 & 1997 & 1998 & 1999 & 2000 & 2001 & 2002 & 2003 \\
\midrule
Estimate ($\widehat{\tau}_{0, t}$) & 953.663 & 834.641 & 40.099 & -465.885 & -670.172 & -803.381 & -1508.747 & -1577.073 & -1737.511 & -2407.415 & -2802.246 & -2936.485 & -3248.218 \\
Lower bound ($\underline{\tau}_{0, t}$) & 202.292 & 177.045 & 8.506 & -1397.656 & -2010.516 & -2410.143 & -4526.242 & -4731.218 & -5212.533 & -7222.244 & -8406.737 & -8809.456 & -9744.655 \\
Upper bound ($\overline{\tau}_{0, t}$) & 2860.988 & 2503.923 & 120.296 & -98.824 & -142.158 & -170.414 & -320.037 & -334.531 & -368.563 & -510.664 & -594.416 & -622.891 & -689.016 \\
\bottomrule
\end{tabular}
}
\vspace{3.5mm}
\caption*{SCPI}
\vspace{-2mm}
\scalebox{0.50}{ 
\begin{tabular}{lrrrrrrrrrrrrr}
\toprule
 & 1991 & 1992 & 1993 & 1994 & 1995 & 1996 & 1997 & 1998 & 1999 & 2000 & 2001 & 2002 & 2003 \\
\midrule
Estimate ($\widehat{\tau}_{0, t}$) & 501.801 & 325.127 & -439.827 & -904.846 & -1109.224 & -1316.447 & -1848.305 & -2119.048 & -2313.993 & -2756.768 & -3075.527 & -3167.449 & -3465.177 \\
Lower bound ($\underline{\tau}_{0, t}$) & 1673.798 & 857.345 & 155.975 & -203.319 & 56.892 & -140.759 & -592.654 & -759.429 & -946.702 & 210.827 & 900.632 & -907.722 & -1186.747 \\
Upper bound ($\overline{\tau}_{0, t}$) & -610.689 & -214.333 & -1032.809 & -1567.552 & -2246.457 & -2198.569 & -2918.746 & -3439.695 & -3480.269 & -4590.658 & -5533.014 & -4855.540 & -5136.086 \\
\bottomrule
\end{tabular}

}
\end{table}

\begin{table}[t]
\caption{Results of treatment effect estimation and conformal inference with the ``Reunification of Germany'' dataset. Estimated treatment effect $\widehat{\tau}_{0, t}$ and $90\%$ confidence interval $(\underline{\tau}_{0, t}, \overline{\tau}_{0, t})$.}
\label{tab:conformal_reunification}

\centering
\caption*{MMSCM ($G = 2$)}
\vspace{-2mm}
\scalebox{0.45}{ 
\begin{tabular}{lrrrrrrrrrrr}
\toprule
 & 1990 & 1991 & 1992 & 1993 & 1994 & 1995 & 1996 & 1997 & 1998 & 1999 & 2000 \\
\midrule
Estimate ($\widehat{\tau}_{0, t}$) & -14.348 & -22.565 & -22.348 & -25.688 & -29.688 & -31.646 & -31.870 & -32.007 & -32.452 & -34.963 & -36.109 \\
Lower bound ($\underline{\tau}_{0, t}$) & -16.957 & -26.668 & -26.411 & -30.359 & -35.086 & -37.400 & -37.665 & -37.826 & -38.353 & -41.320 & -42.674 \\
Upper bound ($\overline{\tau}_{0, t}$) & 5.073 & 7.977 & 7.901 & 9.082 & 10.496 & 11.188 & 11.267 & 11.315 & 11.473 & 12.361 & 12.766 \\
\bottomrule
\end{tabular}
}
\vspace{3.5mm}
\caption*{MMSCM ($G = 3$)}
\vspace{-2mm}
\scalebox{0.45}{ 
\begin{tabular}{lrrrrrrrrrrr}
\toprule
 & 1990 & 1991 & 1992 & 1993 & 1994 & 1995 & 1996 & 1997 & 1998 & 1999 & 2000 \\
\midrule
Estimate ($\widehat{\tau}_{0, t}$) & -14.162 & -22.355 & -22.181 & -25.662 & -29.757 & -31.655 & -31.749 & -32.075 & -32.421 & -34.761 & -35.743 \\
Lower bound ($\underline{\tau}_{0, t}$) & -19.598 & -30.936 & -30.695 & -35.512 & -41.179 & -43.806 & -43.936 & -44.387 & -44.865 & -48.103 & -49.462 \\
Upper bound ($\overline{\tau}_{0, t}$) & 5.007 & 7.903 & 7.842 & 9.072 & 10.520 & 11.191 & 11.225 & 11.340 & 11.462 & 12.289 & 12.636 \\
\bottomrule
\end{tabular}
}
\vspace{3.5mm}
\caption*{MMSCM ($G = 5$)}
\vspace{-2mm}
\scalebox{0.45}{ 
\begin{tabular}{lrrrrrrrrrrr}
\toprule
 & 1990 & 1991 & 1992 & 1993 & 1994 & 1995 & 1996 & 1997 & 1998 & 1999 & 2000 \\
\midrule
Estimate ($\widehat{\tau}_{0, t}$) & -16.070 & -25.074 & -25.344 & -28.937 & -33.229 & -35.489 & -35.507 & -36.330 & -36.699 & -38.613 & -39.638 \\
Lower bound ($\underline{\tau}_{0, t}$) & -20.290 & -31.659 & -32.000 & -36.536 & -41.955 & -44.810 & -44.832 & -45.871 & -46.337 & -48.753 & -50.048 \\
Upper bound ($\overline{\tau}_{0, t}$) & 5.681 & 8.865 & 8.960 & 10.230 & 11.748 & 12.547 & 12.553 & 12.844 & 12.974 & 13.651 & 14.014 \\
\bottomrule
\end{tabular}
}
\vspace{3.5mm}
\caption*{MMSCM ($G = 10$)}
\vspace{-2mm}
\scalebox{0.45}{ 
\begin{tabular}{lrrrrrrrrrrr}
\toprule
 & 1990 & 1991 & 1992 & 1993 & 1994 & 1995 & 1996 & 1997 & 1998 & 1999 & 2000 \\
\midrule
Estimate ($\widehat{\tau}_{0, t}$) & -16.077 & -25.082 & -25.347 & -28.945 & -33.232 & -35.491 & -35.514 & -36.343 & -36.716 & -38.621 & -39.645 \\
Lower bound ($\underline{\tau}_{0, t}$) & -24.196 & -37.749 & -38.148 & -43.563 & -50.016 & -53.416 & -53.450 & -54.698 & -55.260 & -58.126 & -59.668 \\
Upper bound ($\overline{\tau}_{0, t}$) & 6.333 & 9.881 & 9.985 & 11.402 & 13.091 & 13.981 & 13.990 & 14.317 & 14.464 & 15.214 & 15.618 \\
\bottomrule
\end{tabular}
}
\vspace{3.5mm}
\caption*{MMSCM ($G = 50$)}
\vspace{-2mm}
\scalebox{0.45}{ 
\begin{tabular}{lrrrrrrrrrrr}
\toprule
 & 1990 & 1991 & 1992 & 1993 & 1994 & 1995 & 1996 & 1997 & 1998 & 1999 & 2000 \\
\midrule
Estimate ($\widehat{\tau}_{0, t}$) & -19.114 & -28.076 & -28.480 & -31.898 & -36.305 & -38.845 & -38.632 & -39.511 & -40.088 & -41.983 & -42.948 \\
Lower bound ($\underline{\tau}_{0, t}$) & -24.906 & -36.584 & -37.111 & -41.564 & -47.306 & -50.616 & -50.339 & -51.484 & -52.236 & -54.705 & -55.963 \\
Upper bound ($\overline{\tau}_{0, t}$) & 6.757 & 9.926 & 10.069 & 11.277 & 12.835 & 13.733 & 13.658 & 13.968 & 14.173 & 14.842 & 15.184 \\
\bottomrule
\end{tabular}
}
\vspace{3.5mm}
\caption*{MMSCM ($G = 100$)}
\vspace{-2mm}
\scalebox{0.45}{ 
\begin{tabular}{lrrrrrrrrrrr}
\toprule
 & 1990 & 1991 & 1992 & 1993 & 1994 & 1995 & 1996 & 1997 & 1998 & 1999 & 2000 \\
\midrule
Estimate ($\widehat{\tau}_{0, t}$) & -21.245 & -29.901 & -30.293 & -33.607 & -37.968 & -40.632 & -40.331 & -41.157 & -41.945 & -43.863 & -44.732 \\
Lower bound ($\underline{\tau}_{0, t}$) & -25.108 & -35.337 & -35.800 & -39.717 & -44.871 & -48.019 & -47.664 & -48.640 & -49.572 & -51.838 & -52.865 \\
Upper bound ($\overline{\tau}_{0, t}$) & 8.369 & 11.779 & 11.933 & 13.239 & 14.957 & 16.006 & 15.888 & 16.213 & 16.524 & 17.279 & 17.622 \\
\bottomrule
\end{tabular}
}
\vspace{3.5mm}
\caption*{Abadie}
\vspace{-2mm}
\scalebox{0.45}{ 
\begin{tabular}{lrrrrrrrrrrr}
\toprule
 & 1990 & 1991 & 1992 & 1993 & 1994 & 1995 & 1996 & 1997 & 1998 & 1999 & 2000 \\
\midrule
Estimate ($\widehat{\tau}_{0, t}$) & -10.684 & -18.818 & -19.078 & -23.707 & -27.769 & -29.526 & -30.058 & -31.087 & -30.746 & -32.984 & -32.686 \\
Lower bound ($\underline{\tau}_{0, t}$) & -25.577 & -45.048 & -45.673 & -56.752 & -66.477 & -70.684 & -71.957 & -74.420 & -73.604 & -78.963 & -78.249 \\
Upper bound ($\overline{\tau}_{0, t}$) & 3.777 & 6.653 & 6.745 & 8.381 & 9.817 & 10.439 & 10.627 & 10.990 & 10.870 & 11.661 & 11.556 \\
\bottomrule
\end{tabular}
}
\vspace{3.5mm}
\caption*{DiSCo}
\vspace{-2mm}
\scalebox{0.45}{ 
\begin{tabular}{lrrrrrrrrrrr}
\toprule
 & 1990 & 1991 & 1992 & 1993 & 1994 & 1995 & 1996 & 1997 & 1998 & 1999 & 2000 \\
\midrule
Estimate ($\widehat{\tau}_{0, t}$) & -10.403 & -18.998 & -19.634 & -24.119 & -29.899 & -30.819 & -31.128 & -32.149 & -32.234 & -34.680 & -34.051 \\
Lower bound ($\underline{\tau}_{0, t}$) & -24.485 & -44.713 & -46.209 & -56.764 & -70.369 & -72.533 & -73.261 & -75.664 & -75.864 & -81.621 & -80.141 \\
Upper bound ($\overline{\tau}_{0, t}$) & 2.837 & 5.181 & 5.355 & 6.578 & 8.154 & 8.405 & 8.489 & 8.768 & 8.791 & 9.458 & 9.287 \\
\bottomrule
\end{tabular}
}
\vspace{3.5mm}
\caption*{SCPI}
\vspace{-2mm}
\scalebox{0.45}{ 
\begin{tabular}{lrrrrrrrrrrr}
\toprule
 & 1990 & 1991 & 1992 & 1993 & 1994 & 1995 & 1996 & 1997 & 1998 & 1999 & 2000 \\
\midrule
Estimate ($\widehat{\tau}_{0, t}$) & -9.659 & -10.875 & -13.460 & -14.301 & -17.633 & -21.859 & -22.760 & -23.583 & -25.117 & -23.206 & -27.228 \\
Lower bound ($\underline{\tau}_{0, t}$) & -3.876 & -4.886 & -7.346 & -8.138 & -8.889 & -13.372 & -16.258 & -14.058 & -15.537 & -13.986 & -16.757 \\
Upper bound ($\overline{\tau}_{0, t}$) & -14.468 & -16.782 & -18.073 & -18.834 & -20.924 & -26.263 & -27.990 & -28.120 & -31.839 & -30.213 & -31.807 \\
\bottomrule
\end{tabular}
}
\end{table}

\end{document}